\documentclass[prb,aps,epsf,showpacs,twocolumn]{revtex4}
\usepackage{times}
\usepackage{graphicx}
\usepackage{float}
\usepackage{latexsym,amsmath,amssymb,bm,euscript}
\usepackage{color}
\usepackage{subfigure}
\usepackage{epstopdf}
\usepackage{multirow}
\usepackage[colorlinks=true,linkcolor=blue,citecolor=blue]{hyperref}

\def\up{\uparrow}
\def\dn{\downarrow}
\def\Hc{{\rm H.c.}}

\begin{document}

\title{Phase diagram of spin-$\frac{1}{2}$ $J_1$-$J_2$ Heisenberg model on honeycomb lattice}
\author{Shou-Shu Gong$^{1}$, D. N. Sheng$^{1}$, Olexei I. Motrunich$^{2}$, Matthew P. A. Fisher$^{3}$}
\affiliation{$^{1}$Department of Physics and Astronomy, California State University Northridge,
\\$^{2}$Department of Physics, California Institute of Technology,
\\$^{3}$Department of Physics, University of California, Santa Barbara.}

\begin{abstract}
We use density matrix renormalization group (DMRG) algorithm to study the phase diagram of the spin-$1/2$ Heisenberg model on honeycomb lattice with first ($J_1$) and second ($J_2$) neighbor antiferromagnetic interactions, where a $Z_2$ spin liquid region has been proposed. By implementing SU(2) symmetry in the DMRG code, we are able to obtain accurate results for long cylinders with width slightly over $15$ lattice spacings and torus up to the size $N = 2 \times 6 \times 6$. With increasing $J_2$, we find a N\'{e}el phase with vanishing spin gap and a plaquette valence-bond (PVB) phase with non-zero spin gap. By extrapolating the square of the staggered magnetic moment $m_{s}^{2}$ on finite-size cylinders to thermodynamic limit, we find the N\'{e}el order vanishing at $J_2/J_1\simeq 0.22$. For $0.25 < J_2/J_1 \leq 0.35$, we find a possible PVB order, which shows a fast growing PVB decay length with increasing system width. For $0.22 < J_2/J_1 < 0.25$, both spin and dimer orders vanish in thermodynamic limit, which is consistent with a possible spin liquid phase. We present calculations of the topological entanglement entropy, compare the DMRG results with the variational Monte Carlo, and discuss possible scenarios in the thermodynamic limit for this region.

\end{abstract}

\maketitle

\section{Introduction}

The quantum spin liquid (SL) is an enigmatic state of matter where a spin system does not develop magnetic order or break lattice symmetries even at zero temperature and instead develops  a topological order.\cite{Nature_464_199}
Besides being important in the context of the frustrated magnetic systems,\cite{FSS} spin liquid physics may hold clues to theoretic understanding of the non-Fermi liquid behavior of the doped Mott materials\cite{RMP_78_17} and the high-$T_c$ superconductivity of the strongly correlated systems.\cite{MRB_8_153}
The simplest SL's are gapped $Z_2$ states and have been explicitly demonstrated to exist in many model systems including quantum dimer models\cite{PRL_86_1881, PRB_64_064422, PRB_66_205104} and the kagome spin model in the easy axis limit.\cite{PRB_65_224412, PRL_94_146805}
Such a SL is characterized by a $Z_2$ topological order,\cite{PRL_66_1773,PRB_44_2664} a ground state degeneracy on topologically non-trivial manifolds,\cite{PRB_44_2664,PRB_40_7387} as well as fractionalized spinon and vison excitations.\cite{PRB_44_2664, PRB_60_1654, PRB_62_7850}
However, the explicit models for SL phases tend to be fairly contrived and not realistic. It has been a long journey searching for the spin liquids in realistic frustrated spin models, particularly with spin rotational symmetry, that are relevant to real magnetic materials.\cite{PRB_38_9335,PRB_64_104406,PRL_91_107001,PRL_95_036403,PRB_72_045105,PRB_74_012407,PRL_98_107204,PRL_99_137207,PRB_77_224413,PRB_78_064422,JACS_131_8313} Experimentally, some frustrated antiferromagnetic materials indeed can resist forming the magnetic order or breaking real space symmetry at very low temperature,\cite{PRL_98_107204,PRL_99_137207,PRB_78_064422,JACS_131_8313}
while the nature of such states remains to be settled.\cite{PRB_38_9335,PRB_64_104406,PRL_91_107001,PRL_95_036403,PRB_72_045105,PRB_77_224413}

\begin{figure}[t]
\includegraphics[width = 1.0\linewidth,clip]{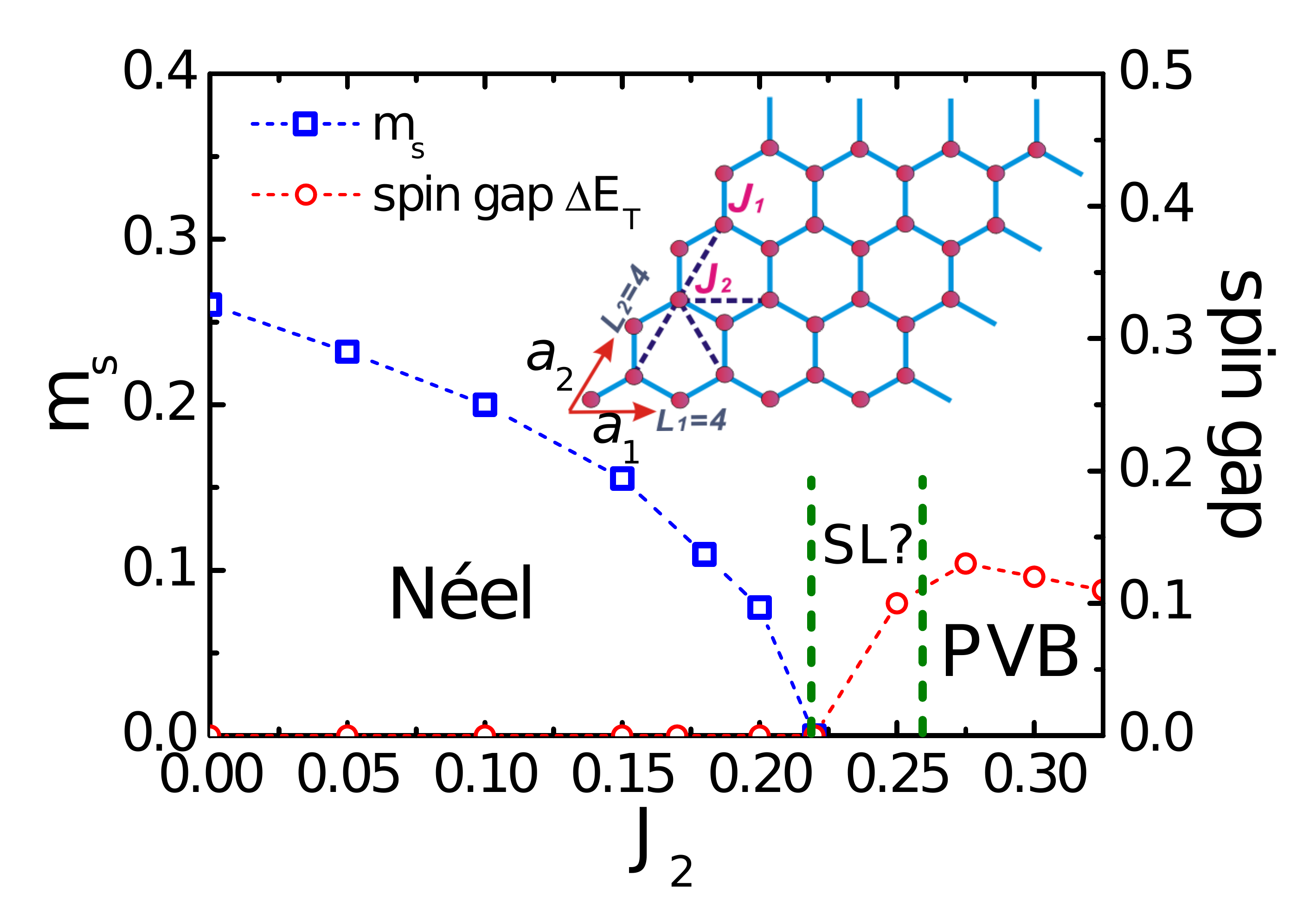}
\caption{Phase diagram of the spin-$\frac{1}{2}$ $J_1$-$J_2$ honeycomb Heisenberg model for $J_2\leq 0.35$ obtained by our SU(2) DMRG studies. With increasing $J_2$, the model has a N\'{e}el phase for $J_2\lesssim 0.22$ and a PVB phase for $0.25\lesssim J_2\lesssim 0.35$. Between these two phases, there is a small region that exhibits no order in our calculations. The main panel shows N\'{e}el order parameter $m_s$ and spin gap $\Delta E_T$. The inset is the sketch of the $J_1$-$J_2$ honeycomb lattice on a $N = 2\times L_1\times L_2$ torus (here with four unit cells, $L_1=L_2=4$, along the two primitive vector directions).} \label{phase}
\end{figure}

Interestingly, large scale DMRG simulations have revealed possible SL phases on kagome\cite{PRL_101_117203,Science_332_1173,PRL_109_067201} and $J_1$-$J_2$ square lattice Heisenberg models.\cite{PRB_86_024424,arxiv_1112_3331}  A recent determinantal quantum Monte-Carlo study has suggested the existence of a spin liquid phase in the half filled Hubbard model on honeycomb lattice;\cite{Nature_464_847} however, a later work appears to contradict this conclusion.\cite{arxiv_1207_1783, AssaadHerbut2013,nonBravais} The related spin models on honeycomb lattice have also attracted intensive attention.\cite{EPJB_20_241,JPCM_23_226006,PRB_84_024406,PRB_81_214419,PRB_83_094506,PRB_82_024419,PRB_84_024420,
PRB_87_024415,PRL_107_087204,PRB_84_014402,PRB_84_014417,PRB_84_094424,PRB_85_060402,JPCM_24_236002,arxiv_1207_1072,arxiv_1208_2416,PRL_110_127203,PRL_110_127205}
Slave-particle approaches\cite{PRB_82_024419,PRB_84_024420,PRB_87_024415} and variational Monte Carlo (VMC) simulations\cite{PRL_107_087204} have proposed a gapped SL in the spin-$\frac{1}{2}$ $J_1$-$J_2$ Heisenberg model and have found relatively low variational energy close to the exact energy obtained from small system exact diagonalization (ED) calculations around $J_2/J_1=0.2$.\cite{PRB_84_024406}
The Hamiltonian of the model is
\begin{equation}
H=J_{1}\sum_{\langle i,j\rangle}\textbf{S}_{i}\cdot\textbf{S}_{j}+J_{2}\sum_{\langle\langle i,j\rangle\rangle}\textbf{S}_{i}\cdot\textbf{S}_{j},
\end{equation}
where the sums $\langle i,j\rangle$ and $\langle\langle i,j\rangle\rangle$ run over all the nearest-neighbor (NN) and the next nearest-neighbor (NNN) bonds, respectively. It has been established that there is a N\'{e}el state on the small $J_2$ side ($J_2\lesssim 0.2J_1$)\cite{JPCM_23_226006,PRB_84_024406,PRB_87_024415,PRB_85_060402,JPCM_24_236002} and a staggered valence-bond (SVB) phase on the large $J_2$ side ($J_2\gtrsim 0.4J_1$).\cite{EPJB_20_241,JPCM_23_226006,PRB_81_214419,PRB_87_024415,PRL_107_087204,PRB_84_014417}
In the intermediate $J_2$ region, the plaquette valence-bond (PVB) state appears strongly in the presence of an additional antiferromagnetic third NN coupling $J_3$.\cite{EPJB_20_241,PRB_84_024406} However, the fate of the quantum state for intermediate $J_2/J_1 \simeq 0.2$ without $J_3$ coupling remains challenging, where competing possibilities include a quantum SL, the PVB state, or a quantum critical point between the N\'{e}el and PVB states.

Very recently, DMRG approach has been applied to study the $J_1$-$J_2$ honeycomb model.\cite{PRL_110_127203,PRL_110_127205} By extrapolating the finite-size spin and dimer orders measured in the bulk of systems with fully open boundaries, Ref.~\onlinecite{PRL_110_127203} finds the N\'{e}el order vanishing at $J_2/J_1\simeq 0.22$, the PVB phase for $0.22\lesssim J_2/J_1\lesssim 0.35$, and the SVB phase for $J_2/J_1\gtrsim 0.35$. Both the transitions are suggested to be continuous and thus indicate the deconfined quantum criticality.\cite{PRB_70_144407}  
On the other hand, Ref.~\onlinecite{PRL_110_127205} systematically measures bulk properties using cylinder systems with open ends, and the authors determine the N\'{e}el order vanishing at $J_2/J_1\simeq 0.26$. For $0.26\lesssim J_2/J_1\lesssim 0.36$, the PVB correlation length grows faster or close to linear with cylinder width, and it is suggested that the system is either quantum critical or has weak PVB order. For $J_2/J_1\gtrsim 0.36$, Ref.~\onlinecite{PRL_110_127205} also finds the SVB phase.  Both works~\onlinecite{PRL_110_127203}~and~\onlinecite{PRL_110_127205} suggest the PVB phase for $0.26\lesssim J_2/J_1 \lesssim 0.35$, but there is still a discrepancy for $0.22 < J_2/J_1 < 0.26$, where a gapped SL had been proposed.\cite{PRB_82_024419, PRB_84_024420, PRB_87_024415, PRL_107_087204}

In this article, we study the $J_1$-$J_2$ Heisenberg model on the honeycomb lattice using the DMRG\cite{PRB_48_10345} with spin rotational SU(2) symmetry\cite{SU2} and the VMC simulations.  We set $J_1$ as energy scale, and lattice spacing between nearest-neighbor sites as length scale. By extrapolating the staggered magnetic moment $m_{s}^{2}$ on cylinder systems with width slightly over $15$ lattice spacings (while the largest sizes are $10$ and $12$ lattice spacings in Refs.~\onlinecite{PRL_110_127203} and \onlinecite{PRL_110_127205}, respectively), we find the N\'{e}el order vanishing at $J_2\simeq 0.22$. To determine the PVB order, we study the width dependence of the PVB correlation length on the cylinder systems, where open boundaries break translational symmetry. We find the PVB correlation length grows strongly with increasing system width for $0.25<J_2\lesssim 0.35$. In the widest cylinders with width larger than $15$ lattice spacings, we observe the long-range PVB order emerging with energy lower than the uniform state. The N\'{e}el and PVB phases are consistent with the gapless and gapped spin excitations extrapolated from the finite-size spin gaps on torus. 

The spin and dimer orders vanish in 2D limit through finite-size scaling for $0.22< J_2\leq 0.25$. To check the possible topological nature of the state, we obtain the topological entanglement entropy (TEE) $\gamma$ by extrapolating the entanglement entropy (EE).\cite{PRL_96_110404,PRL_96_110405,NP_8_902}
It is found that $\gamma\simeq 0.51$ for $0.22<J_2\leq 0.25$. For $J_2=0.3$, $\gamma\simeq 0.66$ is close to the TEE value of $\ln 2$ of $Z_2$ SL, even though the system has PVB order; this indicates that the TEE is not a conclusive measure on our system sizes.

We also compare the spin and dimer correlations at $J_2=0.25$ on the $N=2\times 6\times 6$ torus with VMC wave functions at different parameters, and find a striking match from a $Z_2$ SL trial wave function. While our finite size results are consistent with a SL phase, we are challenged by the fact that spin liquid is not likely to have a continuous transition to N\'{e}el phase,\cite{IJMP_5_219} making it also possible that the system has a N\'{e}el-PVB deconfined quantum critical point with larger length scale beyond our system length.

By employing SU(2) symmetry in DMRG, we can get access to larger system sizes with high accuracy, which is essential for distinguishing a SL from competing weakly ordered states.
For cylinder systems with open edge, the U(1) DMRG is usually limited to the system width of $12$ lattice spacings by keeping $6000\sim 8000$ states.\cite{PRL_110_127203,PRL_110_127205} In our calculations, we study the cylinder systems with width more than $15$ lattice spacings by keeping up to $24000$ states to obtain the converged results. With the SU(2)-symmetric implementation, we can also study the torus system up to the size $2\times 6\times 6$ by keeping more than $40000$ states. The truncation error is controlled below $10^{-6}$ in most cases, which gives well converged  results.

\begin{figure}[t]
\includegraphics[width = 1.0\linewidth,clip]{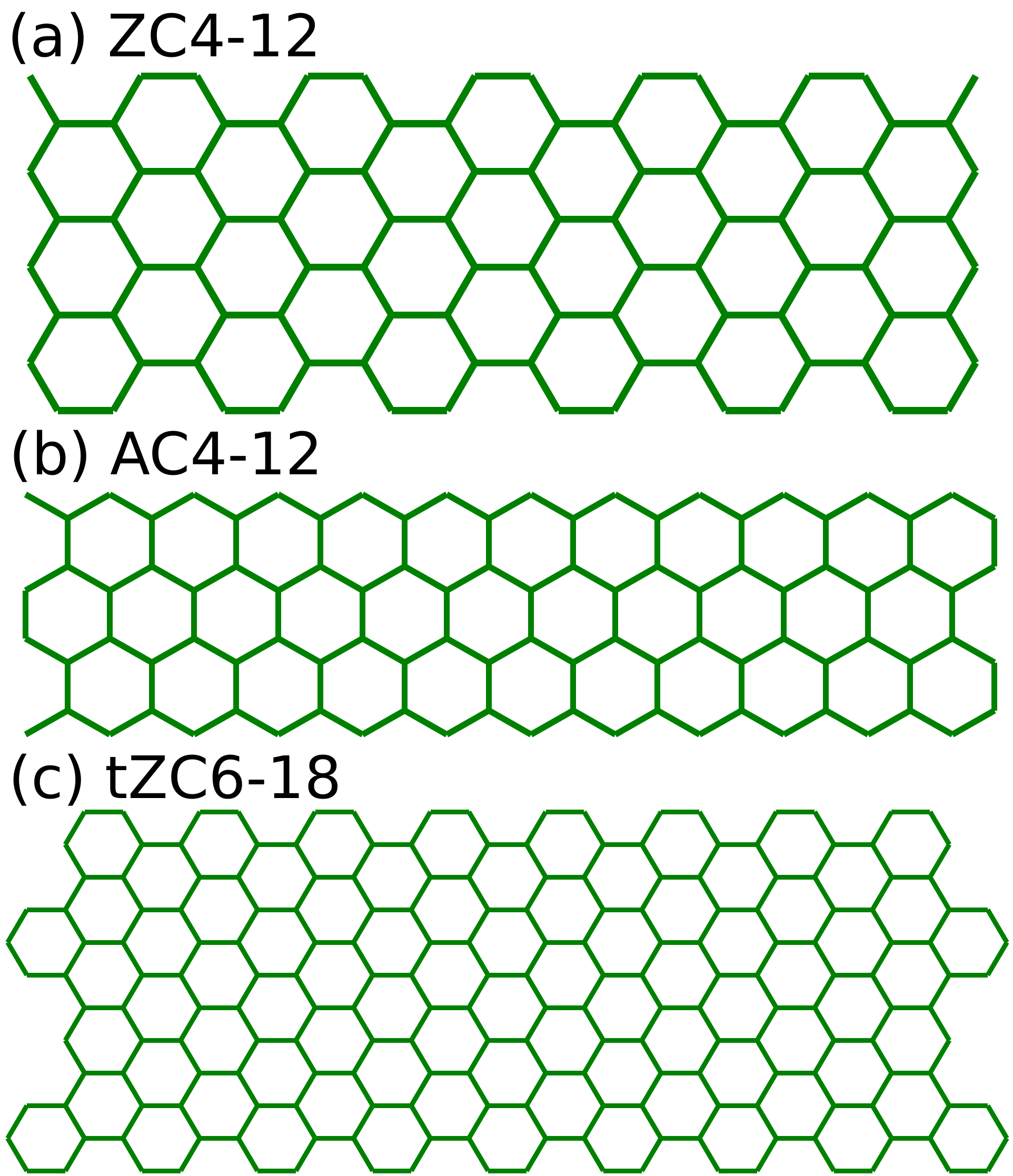}
\caption{Cylinders used in DMRG calculations. (a) ZC4-12 cylinder with zigzag open edges. It has $4$ unit cells along the zigzag direction ($W_y=4\sqrt{3}$) and $12$ columns along the axis direction. (b) AC4-12 cylinder with armchair open edges. It has $4$ vertical bonds along the armchair direction ($W_y=6$) and $12$ armchair columns along the axis direction. (c) Trimmed ZC cylinder tZC6-18 with trimmed zigzag edges. It has $6$ unit cells along the zigzag direction ($W_y=6\sqrt{3}$) and $18$ columns along the axis direction.} 
\label{lattice}
\end{figure}

\begin{figure}[t]
\includegraphics[width = 1.0\linewidth,clip]{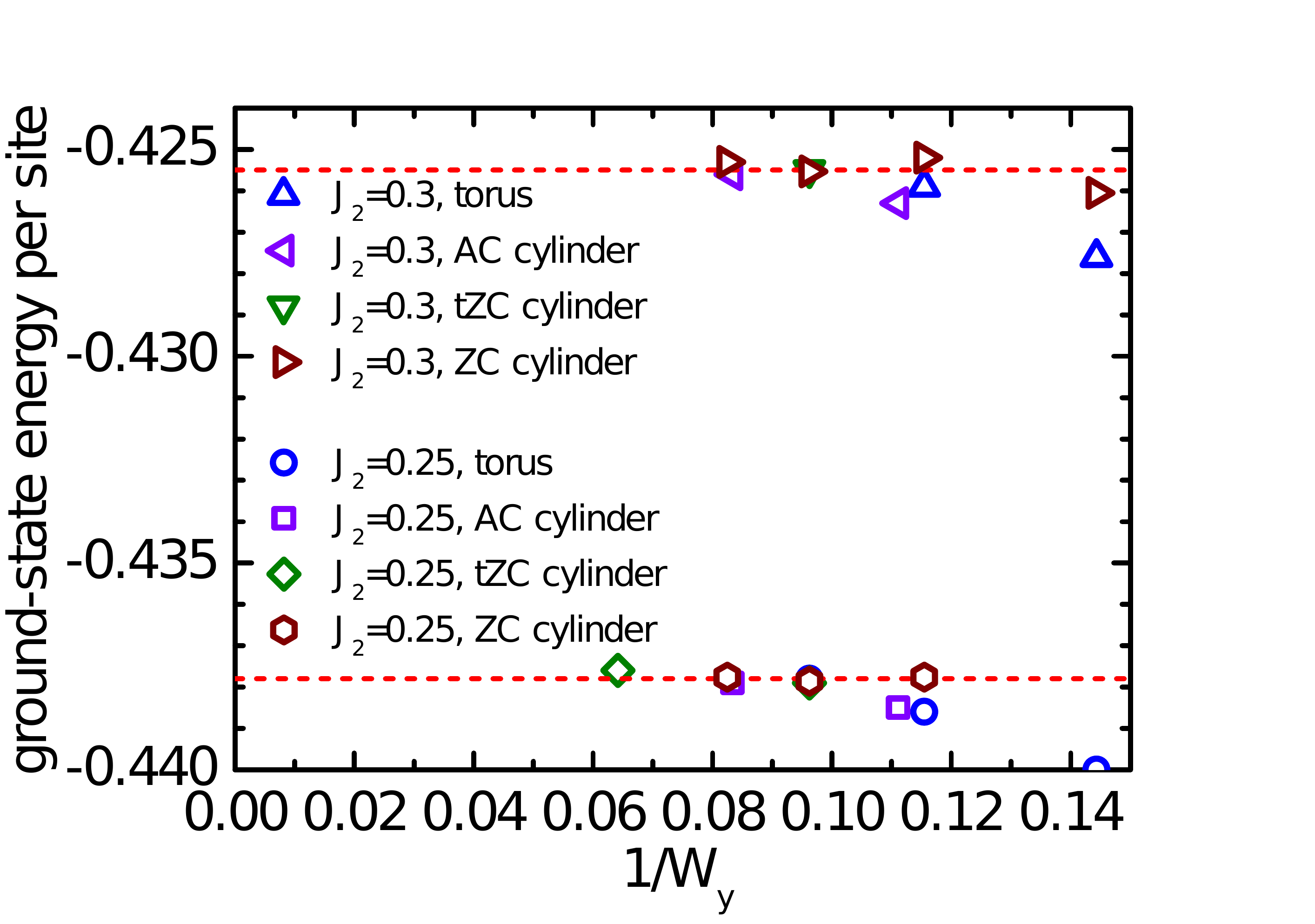}
\caption{Circumference dependence of the ground-state energy per site for $J_2=0.25$ and $0.3$ on torus, AC, tZC, and ZC cylinders. Each data is obtained by keeping the optimal states to our computation limit. The truncation errors are below $1\times 10^{-6}$ except for the largest size at $J_2=0.25$ (tZC9 cylinder with $W_y\simeq 15.588$), where the error is about $5\times 10^{-6}$. The dashed lines indicate the extrapolations of the energies, which give $-0.4378$ and $-0.4255$ for $J_2=0.25$ and $0.3$, respectively.}\label{energy}
\end{figure}

We study the model on both torus and cylinder. The torus geometry is denoted as $N=2\times L_1\times L_2$, where $L_1$ and $L_2$ are the number of unit cells along the two primitive vector directions (the inset of Fig.~\ref{phase} shows the $N=2\times 4\times 4$ torus). For cylinder geometry, we study the systems with three different boundaries.
The first cylinder [Fig.~\ref{lattice}(a)] has the zigzag open edges and is denoted as ZC$m$-$n$ cylinder, where $m$ is the number of two-site unit cells along the column and $n$ is the number of columns along the axis direction. The ZC$m$-$n$ cylinder is equivalent to the XC$2m$ cylinder in Ref.~\onlinecite{PRL_110_127205}. DMRG calculations in our studied region give the uniform states without translational symmetry breaking in ZC cylinder. To induce the PVB order, we can change the couplings of some edge bonds to introduce pinning force.
The second cylinder AC$m$-$n$ [Fig.~\ref{lattice}(b)] has the armchair open edges, where $m$ is number of unit cells in the column direction and must be even to form the periodic boundary condition in column direction; this system is equivalent to the YC$m$ cylinder in Ref.~\onlinecite{PRL_110_127205}. AC cylinder accommodates both the PVB and SVB orders, and
its edges can also select among degenerate states within each order.
The third cylinder is obtained by trimming the three neighbor sites per six sites along the edges on the ZC cylinder to make the lattice strongly select particular PVB state. This system is denoted as tZC$m$-$n$ cylinder, where $m$ must be multiple of $3$ to form the periodic boundary condition in column direction, and is shown in Fig.~\ref{lattice}(c).  In our DMRG calculations, we use ZC cylinder to study $m_{s}^{2}$ to determine the vanishing of N\'{e}el order, and we use all three cylinders to study the PVB order.  To demonstrate the results of AC and tZC (ZC) cylinders together, we also use the circumference $W_y$ to denote the geometrical width of cylinders in units of nearest-neighbor spacing. On AC$m$ and tZC$m$ (ZC$m$) cylinders, the circumferences are $W_y=1.5\times m$ and $\sqrt{3}\times m$, respectively.

To check the accuracy of our computations, we present the circumference dependence of the ground-state energy per site on torus, AC, tZC, and ZC cylinders for both $J_2=0.25$ and $0.3$ in Fig.~\ref{energy}. We obtained the data by keeping the optimal states to our computation limit; the truncation errors are below $1\times 10^{-6}$ except for the largest size (tZC9 cylinder with $W_y\simeq 15.588$) at $J_2=0.25$, where the truncation error is about $5\times 10^{-6}$. To eliminate boundary effects, we calculate the bulk energy on cylinder by subtracting the energies of two samples with different lengths.\cite{ARCMP_3_111} By extrapolating the energies in Fig.~\ref{energy}, we estimate $-0.4378$ and $-0.4255$ as the thermodynamic limit ground-state energies for $J_2=0.25$ and $0.3$, respectively. The latter value is consistent with the result in Ref.~\onlinecite{PRL_110_127205}.

The remainder of the paper is organized as follows. In Sec.~\uppercase\expandafter{\romannumeral2}, we calculate the square of the staggered magnetic moment, $m_{s}^{2}$, on ZC cylinder for various $J_2$ and extrapolate the finite-size data to thermodynamic limit to estimate the N\'{e}el order. In Sec.~\uppercase\expandafter{\romannumeral3}, we study the PVB order on AC, ZC, and tZC cylinders from the N\'{e}el to the intermediate region. In Sec.~\uppercase\expandafter{\romannumeral4}, we obtain the spin gaps on finite-size torus and extrapolate to thermodynamic limit. We study the EE and TEE in Sec.~\uppercase\expandafter{\romannumeral5} to check the possible topological nature for the intermediate region. In Sec.~\uppercase\expandafter{\romannumeral6}, we compare the DMRG results with the variational wave functions based on slave-fermion approach, while in Sec.~\uppercase\expandafter{\romannumeral7} we discuss our results and summarize.
In Appendix~\ref{app:SBVMC}, we also present variational results using Schwinger Boson construction.

\section{Magnetic order}

\begin{figure}[t]
\includegraphics[width = 0.9\linewidth,clip]{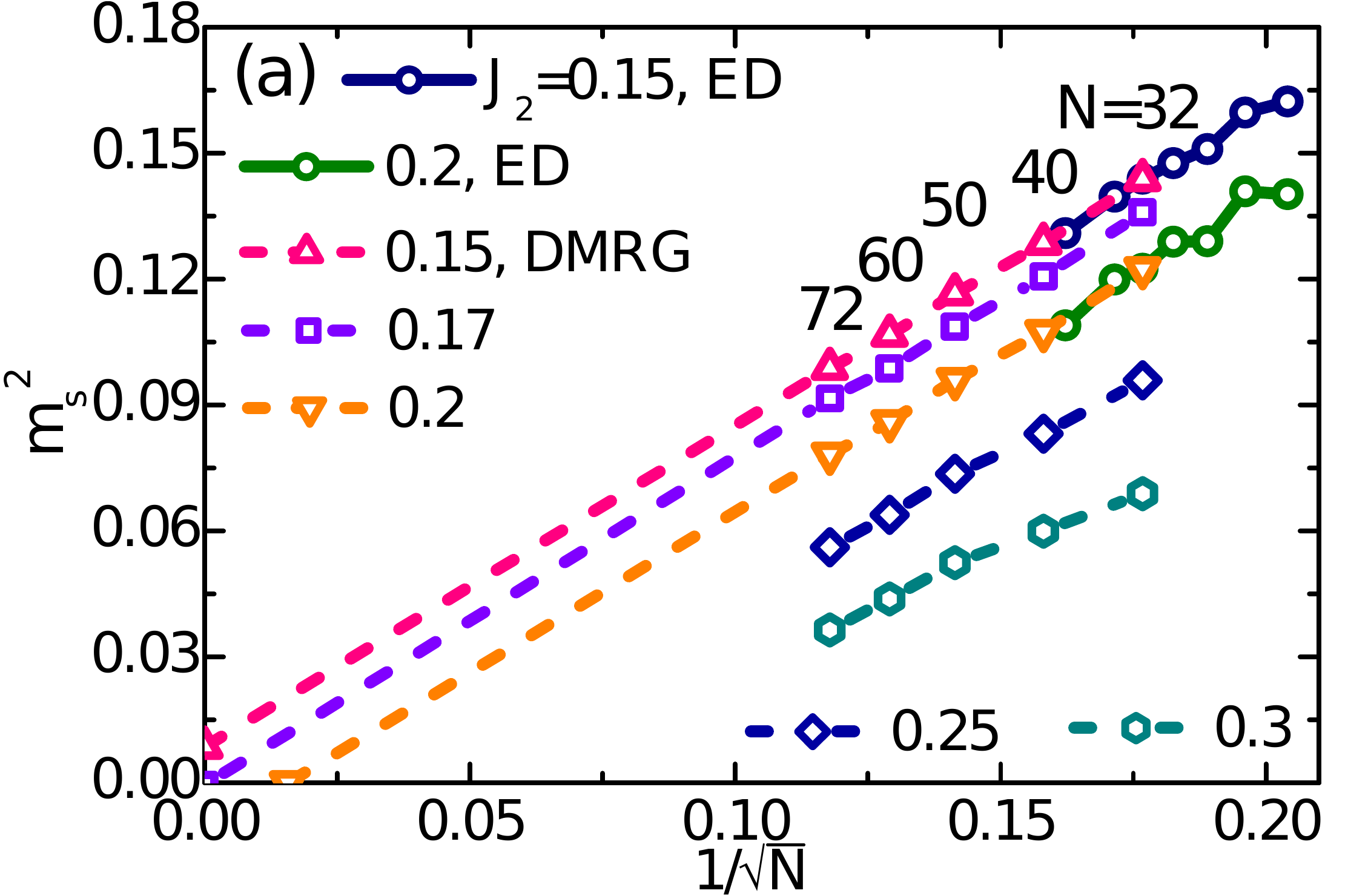}
\includegraphics[width = 1.0\linewidth,clip]{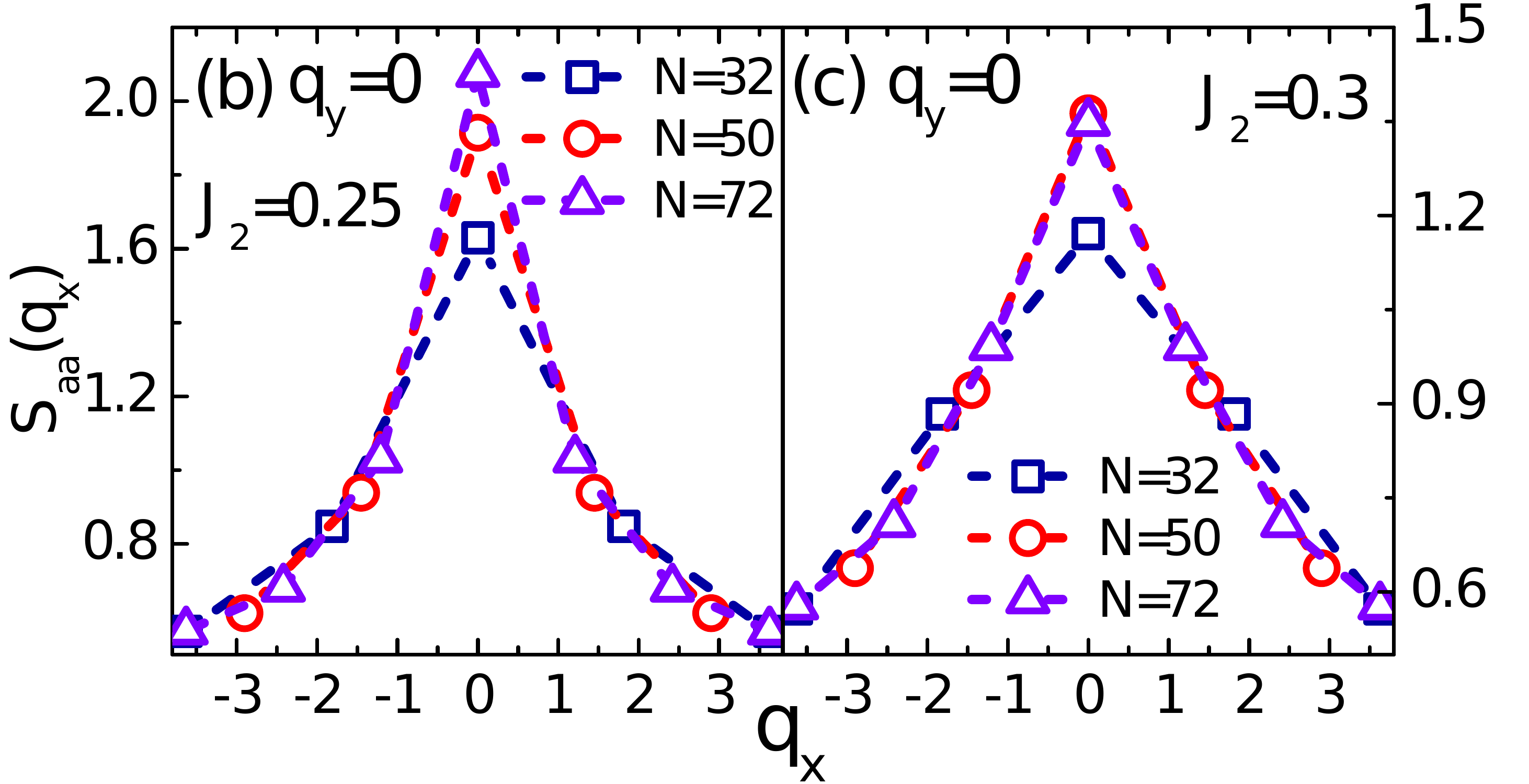}
\includegraphics[width = 1.0\linewidth,clip]{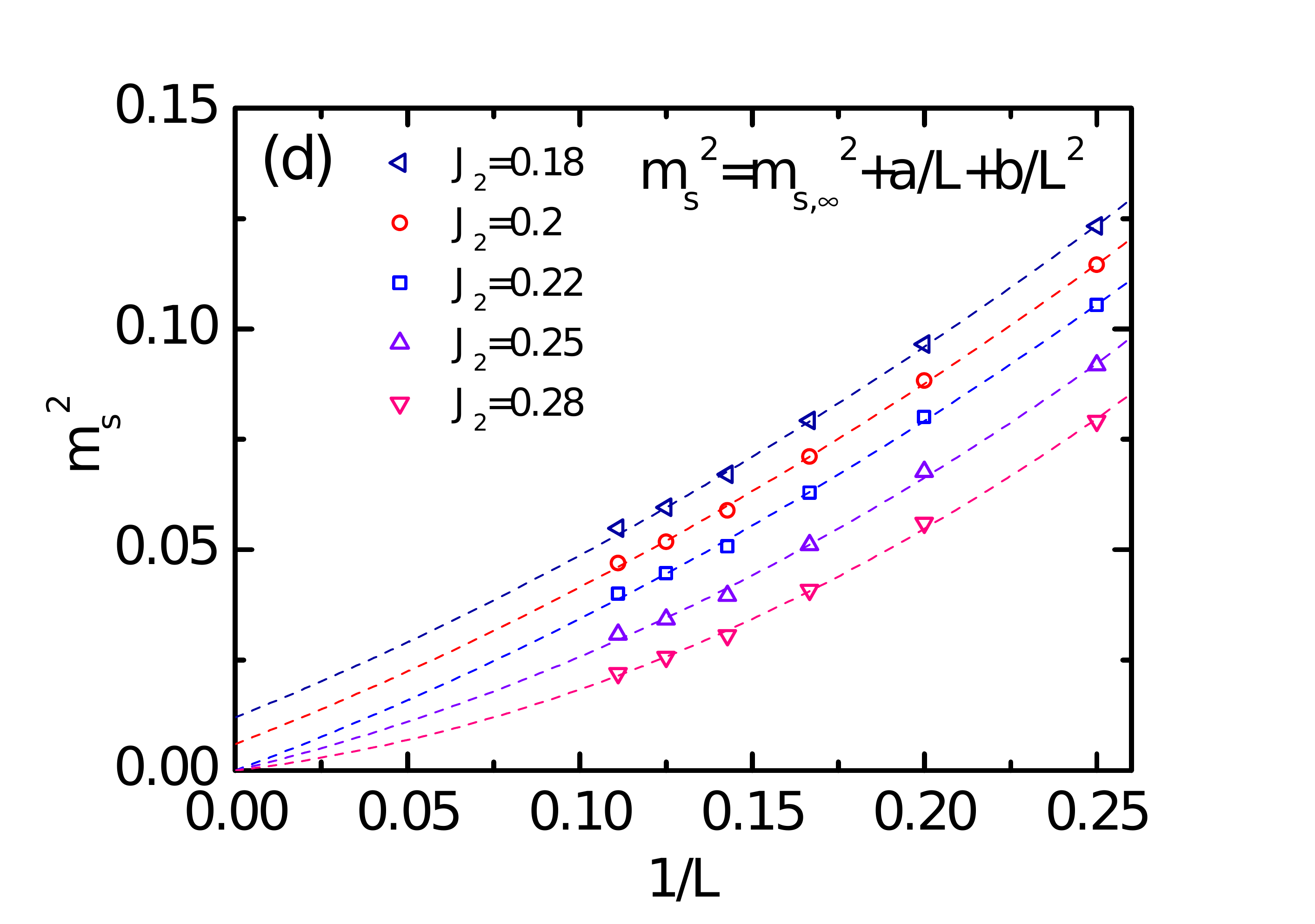}
\caption{(a) $m_{s}^{2}$ plotted vs $1/\sqrt{N}$ for the torus clusters $N=2\times 4\times 4$, $2\times 5\times 4$, $2\times 5\times 5$, $2\times 6\times 5$, and $2\times 6\times 6$. The ED data is from Ref.~\onlinecite{PRB_84_024406}. (b),(c) Size dependence of same-sublattice spin structure factor obtained on torus for $J_2=0.25$ and $0.3$, respectively. The system sizes are $N=2\times 4\times 4$, $2\times 5\times 5$, and $2\times 6\times 6$. (d) $m_s^2$ plotted vs $1/L$ for the ZC$L$-$2L$ cylinder with $L=4,5,6,7,8,9$. Here $m_s^2$ is obtained from $N/2$ spins in the middle part of the sample.}
\label{m}
\end{figure}

The N\'{e}el order on honeycomb lattice is described by the staggered magnetic moment
$m_{s}^{2}=\langle(\sum_{i}(-1)^{i}\mathbf{S}_{i})^{2}\rangle/N^{2}$.\cite{PRB_84_024406}
We obtain the staggered magnetic moment by calculating the spin-spin correlation functions on both torus and cylinder.
In Fig.~\ref{m}(a), we plot DMRG data on torus together with smaller size ED data\cite{PRB_84_024406} for $m_s^2$ at various system sizes $N$ and a few $J_2$ closer to the possible transition point (around $0.2$ identified by ED\cite{PRB_84_024406}) as a function of $1/\sqrt{N}$. The leading $1/\sqrt{N}$ correction of the finite-size scaling is well satisfied in these clusters\cite{PRB_39_2608} through the good straight line fitting to all data points with $J_2 \leq 0.17$.

For $J_2=0.3$ deep in the intermediate region, the spin correlations decay exponentially in real space.  We can also see this by examining the structure factor of the spin correlations between the sites in the same sublattice $S_{aa}(\vec q)$ (for sublattice $A$)
\begin{equation}
S_{aa}(\vec q)=\frac{1}{L_1 L_2}\sum_{i \in A, j \in A}\left\langle \mathbf{S}_{i}\cdot\mathbf{S}_{j}\right\rangle e^{i\vec q\cdot (\vec{r}_i-\vec{r}_j)}. \label{saa}
\end{equation}
In Figs.~\ref{m}(b) and \ref{m}(c), we present $S_{aa}(\vec q)$ for $J_2=0.25$ and $0.3$ obtained on torus. For each system, we see a peak at momentum ${\vec q} = 0$ corresponding to N\'{e}el-like spin correlation in real space. For $J_2=0.25$ the peak is still growing but more slowly than a linearly in $N$, while for $J_2=0.3$ the peak is saturating already at the size $N=2\times 5\times 5$, which are consistent with the vanishing of $m_s$ in the thermodynamic limit. The same behavior is also obtained for the structure factor of the spin correlations between the $A$ and $B$ sublattices. Such torus data on our sizes therefore further support that the N\'{e}el order is absent at least for $J_2 \geq 0.25$.

However, the torus boundary condition limits the system size in DMRG calculations due to larger truncation error for the same number of states kept.\cite{PRB_48_10345} Therefore, we extend the system size by studying cylinder system. We choose the ZC$L$-$2L$ cylinder with system size $N=2\times 2L\times L$. The magnetic moment $m_s$ is obtained from the spin-spin correlations of the $N/2$ sites in the middle of sample, which effectively reduces the boundary effect.\cite{PRB_86_024424,PRL_99_127004} We calculate $m_s^2$ for samples with $L=4$ to $9$ and show the results in Fig.~\ref{m}(d). The finite-size $m_s^2$ at $L = 4, 5, 6$ are close to the results on torus and their extrapolations are consistent with those in Fig. \ref{m}(a). However, on larger sizes the results deviate from the straight line extrapolations of the small-size data. We therefore fit the data using the formula $m_s^2 = m_{s,\infty}^2 + a/L + b/L^2$.  From the best fits we estimate that the N\'{e}el order vanishes at $J_2\simeq 0.22$. This observation is consistent with the DMRG result in Ref.~\onlinecite{PRL_110_127203}, where the finite-size $m_s^2$ are obtained on two different system samples with fully open boundaries up to the size $L=6$.
On the other hand, in Ref.~\onlinecite{PRL_110_127205} the authors estimate the 2D magnetic order parameter by applying a staggered field at the open ends of cylinder with optimal aspect ratio and measuring the local $\langle S^{z}\rangle$ at the center of the sample. They determine that the transition occurs at $J_2\simeq 0.26$. While both methods of extrapolating $m_{s,\infty}$ are standard, they are limited by the reachable system size, and therefore the exact vanishing point of N\'{e}el order might still be an open question.

\section{Plaquette valence bond order}

\begin{figure}[t]
\includegraphics[width = 0.49\linewidth,clip]{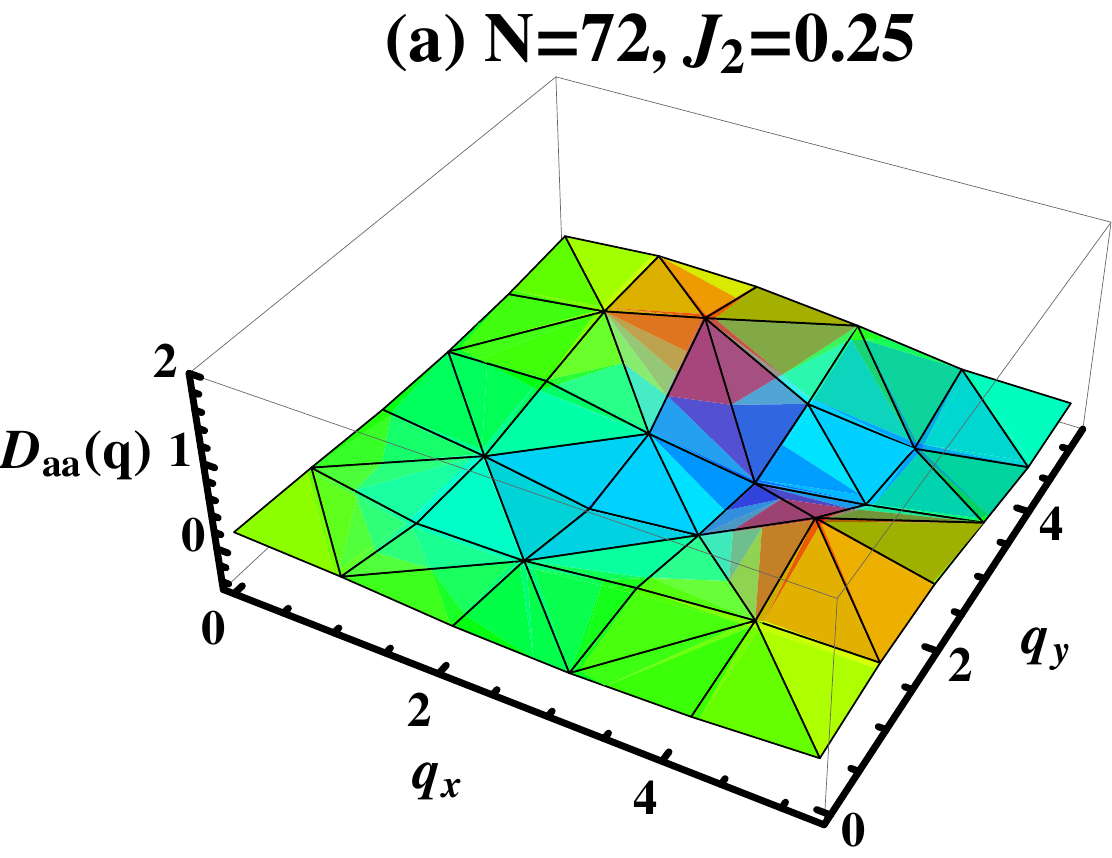}
\includegraphics[width = 0.49\linewidth,clip]{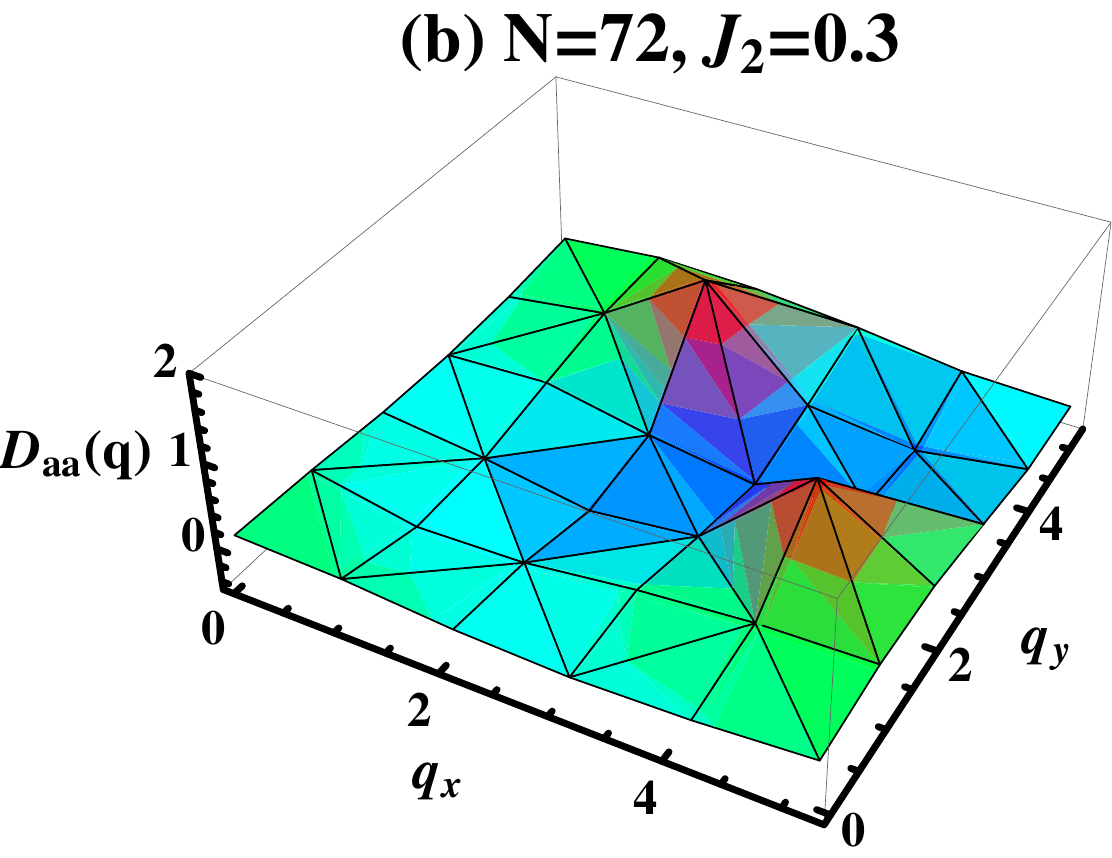}
\caption{Structure factor of dimer-dimer correlation functions on the $N=2\times 6\times 6$ torus for (a) $J_2=0.25$ and (b) $J_2=0.3$.} \label{sq_dimer}
\end{figure}

To detect or exclude the possible VBS order in the intermediate region, we can study the dimer-dimer correlation function
\begin{equation}
C_{(i,j),(k,l)} = 4 \left[ \langle(\mathbf{S}_i\cdot\mathbf{S}_j)(\mathbf{S}_k\cdot\mathbf{S}_l)\rangle-\langle\mathbf{S}_i\cdot\mathbf{S}_j\rangle\langle\mathbf{S}_k\cdot\mathbf{S}_l\rangle \right] \label{c}
\end{equation}
in the system without lattice symmetry breaking, where $(i,j)$ and $(k,l)$ are NN bonds. We can also examine the corresponding structure factors defined as the Fourier transform of the dimer correlations, and here we consider the dimers oriented in the same direction:
\begin{equation}
D_{aa}(\vec{q})=\frac{1}{L_1 L_2}\sum_{(i,j),(k,l)}C_{(i,j),(k,l)}e^{i\vec{q}\cdot(\vec{r}_{(i,j)}-\vec{r}_{(k,l)})}.\label{daa}
\end{equation}
On torus system, our DMRG calculations obtain the ground states without lattice symmetry breaking. Therefore, we can study the dimer-dimer correlations. To accommodate the PVB order on torus, both $L_1$ and $L_2$ must be multiples of $3$. In Fig.~\ref{sq_dimer}, we present the dimer structure factor for $J_2=0.25$ and $0.3$ on the $2\times 6\times 6$ torus (this size accommodates both the PVB and SVB orders). For these $J_2$ couplings, the dimer structure factor has two weak peaks at ${\bf q} = (2\pi/3, 4\pi/3)$ and $(4\pi/3, 2\pi/3)$ that are consistent with the possible PVB pattern.  However, it is not clear if the long-range PVB order will form in the large system limit. The absence of peak at ${\bf q} = {\bf 0}$ indicates the absence of SVB order.

To study the PVB order on larger system sizes, we make the DMRG calculations on cylinder systems. An effective method to detect dimer order on cylinder system is proposed in the quantum Monte Carlo study of the J-Q model on the square lattice\cite{PRB_85_134407} (this model has a transition from the N\'{e}el to the columnar dimer phase with changing Q coupling) and the DMRG study of the $J_1$-$J_2$ Heisenberg model on the square lattice.\cite{PRB_86_024424} The idea of the method is to study the width dependence of the decay length of the dimer texture induced near a boundary. For the system without dimer order in the 2D limit, the dimer decay length might increase with growing width but should saturate in the thermodynamic limit, while for the system with dimer order, it will diverge. The DMRG calculations on cylinder could obtain the ground state with lattice symmetry breaking by making the lattice compatible with the possible dimer order. Thus, one could define the local dimer order parameter and measure the decay of the dimer order from boundary to bulk, from which we can estimate a decay length.
To determine the PVB order on honeycomb lattice, we study the width dependence of the dimer decay length on the AC, tZC, and ZC cylinders. On these systems, the local PVB order parameter decays exponentially from boundary to bulk, from which we can estimate the dimer decay length $\xi_{P}$ and investigate its dependence on the cylinder width. The details are discussed below.

\subsection{PVB order on AC cylinder}

\begin{figure}[t]
\includegraphics[width = 1.0\linewidth,clip]{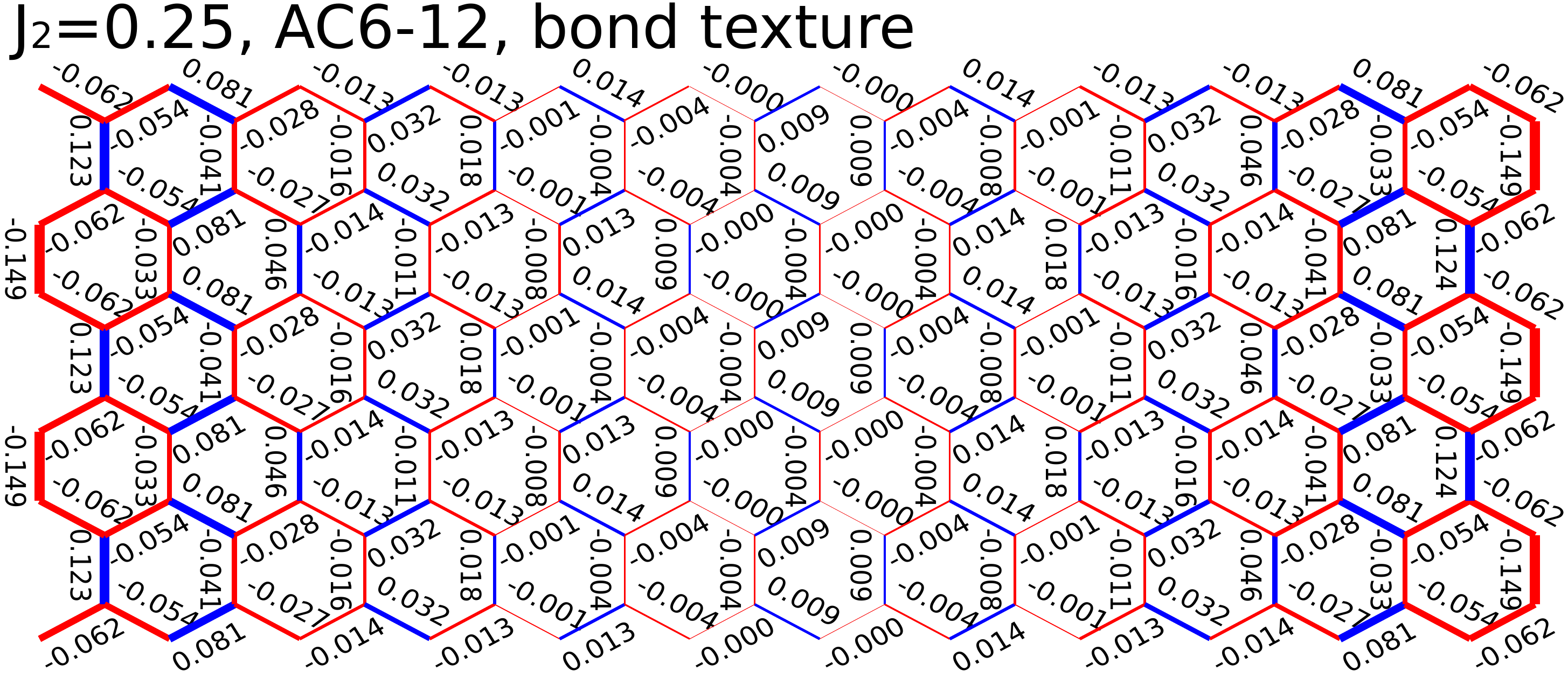}
\caption{The PVB bond texture $B_{i,j}$ for $J_2=0.25$ on an AC6-12 cylinder. The red bonds with negative values have lower NN bond energies. Thus, the red hexagons with negative bonds could indicate the ``resonating plaquettes.''} \label{YC_PVB}
\end{figure}

The AC cylinder with the armchair open boundaries accommodates the PVB order and can select a unique state where the hexagons on the open edges could form ``resonating plaquettes'' (the red hexagons with negative numbers in Fig.~\ref{YC_PVB}). The induced local orders can be identified by the distribution of the subtracted NN bond energy, i.e.\ bond texture defined as
\begin{equation}
B_{i,j} = \langle \textbf{S}_i\cdot \textbf{S}_j\rangle - e_\alpha ~,
\end{equation}
where $e_\alpha$ ($\alpha=1,2,3$) is the average of the NN bond energies in the given bond direction $\alpha$ evaluated in the middle half of system.  Figure~\ref{YC_PVB} shows the bond textures $B_{i,j}$ for $J_2=0.25$ on an AC6-12 cylinder.  The red bonds have lower NN bond energies, and the red hexagons could indicate the ``resonating plaquettes.''  The bond textures decay from boundary to bulk.

\begin{figure}[t]
\includegraphics[width = 1.0\linewidth,clip]{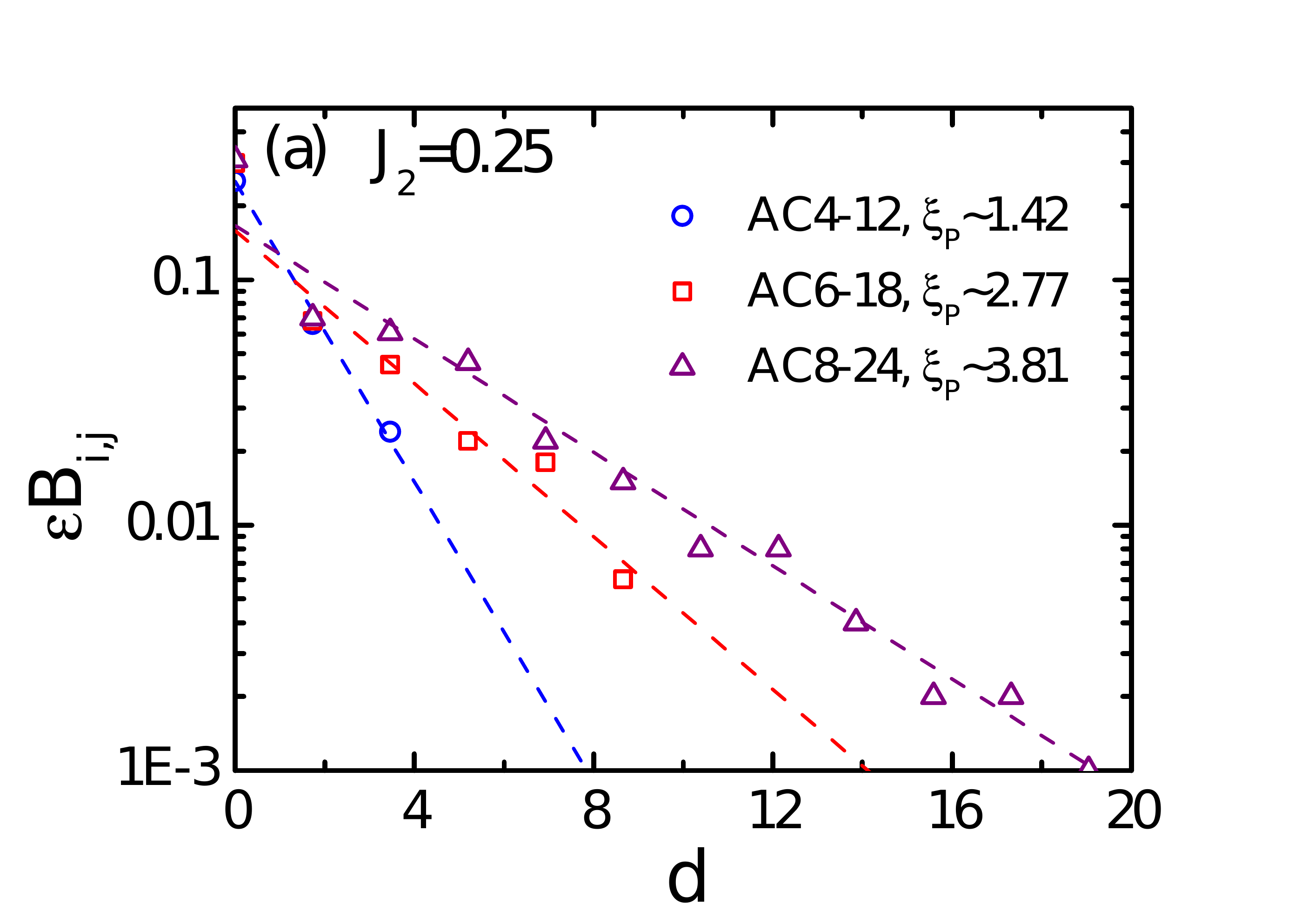}
\includegraphics[width = 1.0\linewidth,clip]{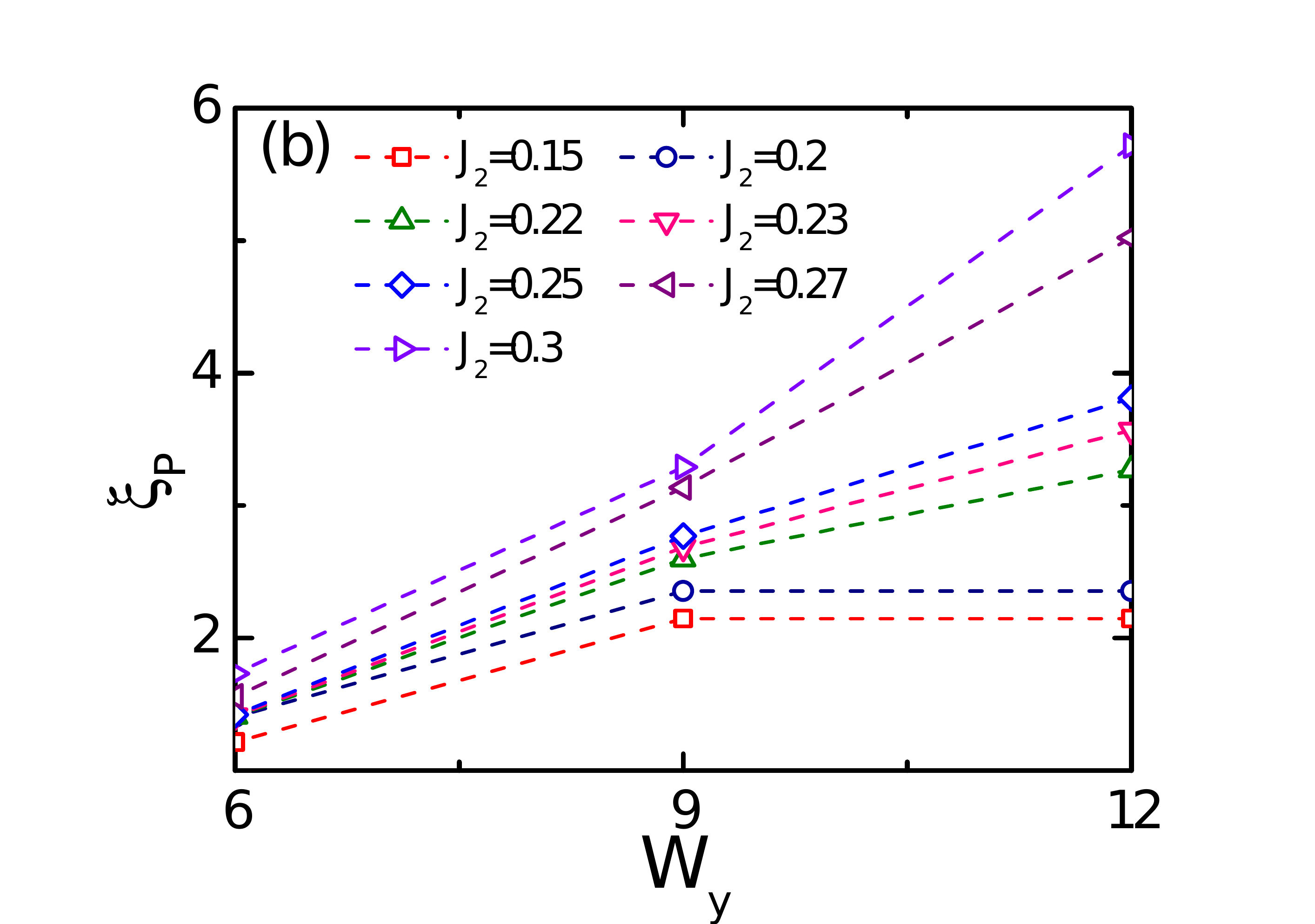}
\caption{(a) Log-linear plot of $\epsilon_{i,j} B_{i,j}$ vs distance from boundary to bulk for $J_2=0.25$ on AC4-12, AC6-18, and AC8-24 cylinders. The decay length $\xi_{P}$ is obtained by fitting the decay behavior of $\epsilon_{i,j} B_{i,j}$. (b) Circumference dependence of the decay length $\xi_{P}$ on AC4-12, AC6-18, and AC8-24 cylinders for various $J_2$ couplings.} \label{YC_decay}
\end{figure}

To describe the decay of bond texture, we multiply the positive and negative $B_{i,j}$ by $\epsilon_{i,j}=1$ and $-2$ respectively (appropriate for the PVB order), and measure the decay of vertical bond $\epsilon_{i,j} B_{i,j}$ from open boundary to bulk. Figure~\ref{YC_decay}(a) is the log-linear plot for the vertical $\epsilon_{i,j} B_{i,j}$ at $J_2=0.25$ on the AC4-12, AC6-18, and AC8-24 cylinders and shows that $\epsilon_{i,j} B_{i,j}$ decays exponentially from edge to bulk. The decay length $\xi_{P}$ increases with increasing cylinder width.  Figure~\ref{YC_decay}(b) shows our study of the circumference dependence of $\xi_{P}$ for various $J_2$ couplings from the N\'{e}el phase to the intermediate region. For $J_2=0.15$ and $0.2$ in the N\'{e}el phase, $\xi_{P}$ is saturated on the AC6 cylinder. For $0.2 < J_2 \leq 0.25$, $\xi_{P}$ grows continuously from AC4-12 to AC8-24, but apparently more slowly than the linear increase, indicating that the dimer decay lengths could be finite in the 2D limit. For $J_2=0.27$ and $0.3$, $\xi_{P}$ grows strongly with increasing width, which implies the diverging decay length in the 2D limit and PVB order.

\begin{figure*}[t]
\includegraphics[width = 0.7\textwidth,clip]{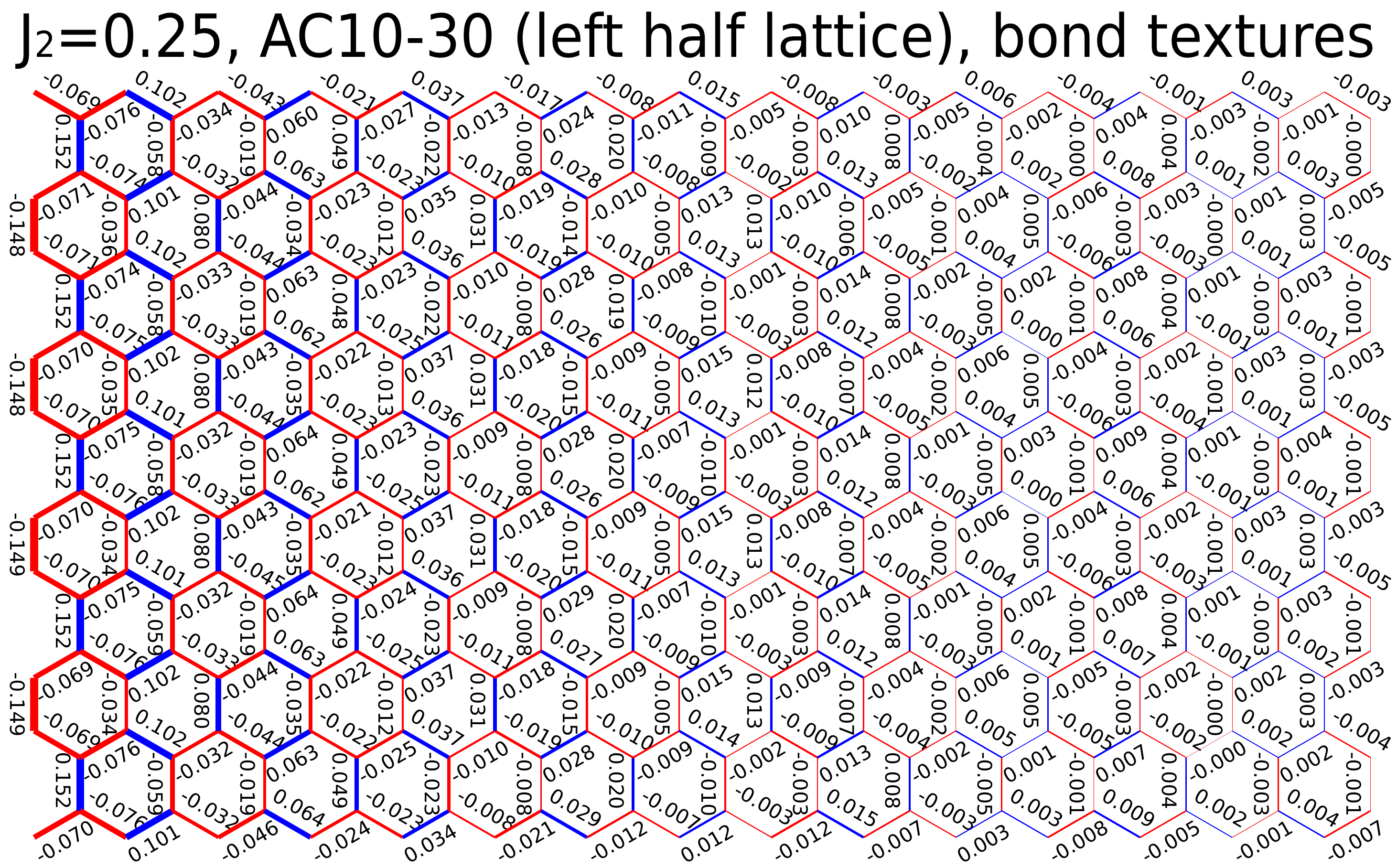}
\caption{The PVB bond texture $B_{i,j}$ for $J_2=0.25$ on an AC10-30 cylinder. For clarity, only the left half of the lattice is shown. For this large size, the truncation error is about $5\times 10^{-6}$ and the bond energies have some small uncertainty of $\pm 0.001$.  At this accuracy, the PVB order vanishes in the middle of the sample.} \label{YC10_025}

\end{figure*}

For $J_2=0.25$ and $0.3$, we also study the PVB order on the AC10-30 cylinder with circumference $W_y=15$.  This is the size limit for AC cylinder in our DMRG calculations, and for such sizes we are no longer sure about the convergence of our measurements of $\xi_P$. We keep more than $20000$ states for DMRG sweeps and obtain the results with the truncation error $5\times 10^{-6}$ for $J_2=0.25$ and $1\times 10^{-5}$ for $J_2=0.3$.  As shown in Fig.~\ref{YC10_025} for $J_2=0.25$, the PVB bond textures are weak in the bulk of system, and the fitted decay length is $\xi_{P}\simeq 5.1$. A linear extrapolation of the $\xi_{P}$ for the AC10 cylinder from the $\xi_{P}$ on the AC6 and AC8 cylinders in Fig.~\ref{YC_decay}(b) would give $\xi_{P}=4.8$. Although our present data is slightly larger than the linear extrapolation result, we expect that $\xi_{P}$ will decrease significantly for this system if we keep even more states, which is beyond our present capability. For example, when we study the AC8-24 cylinder at $J_2=0.25$, our fitted $\xi_{P}$ decreases from $4.7$ to $3.8$ when we increase the number of states kept from $4000$ to $10000$ U(1) equivalent states [the latter number is shown in Fig.~\ref{YC_decay}(b)]. From the present data, we tentatively conclude that $J_2 = 0.25$ does not have PVB order in the 2D limit.
On the other hand, from similar visualization of the bond texture for $J_2=0.3$ (not shown), we observe a long-range PVB order that is consistent with the strong growth of $\xi_{P}$ in Fig.~\ref{YC_decay}(b).

\subsection{PVB order on the trimmed ZC cylinder}

\begin{figure}[t]
\includegraphics[width = 1.0\linewidth,clip]{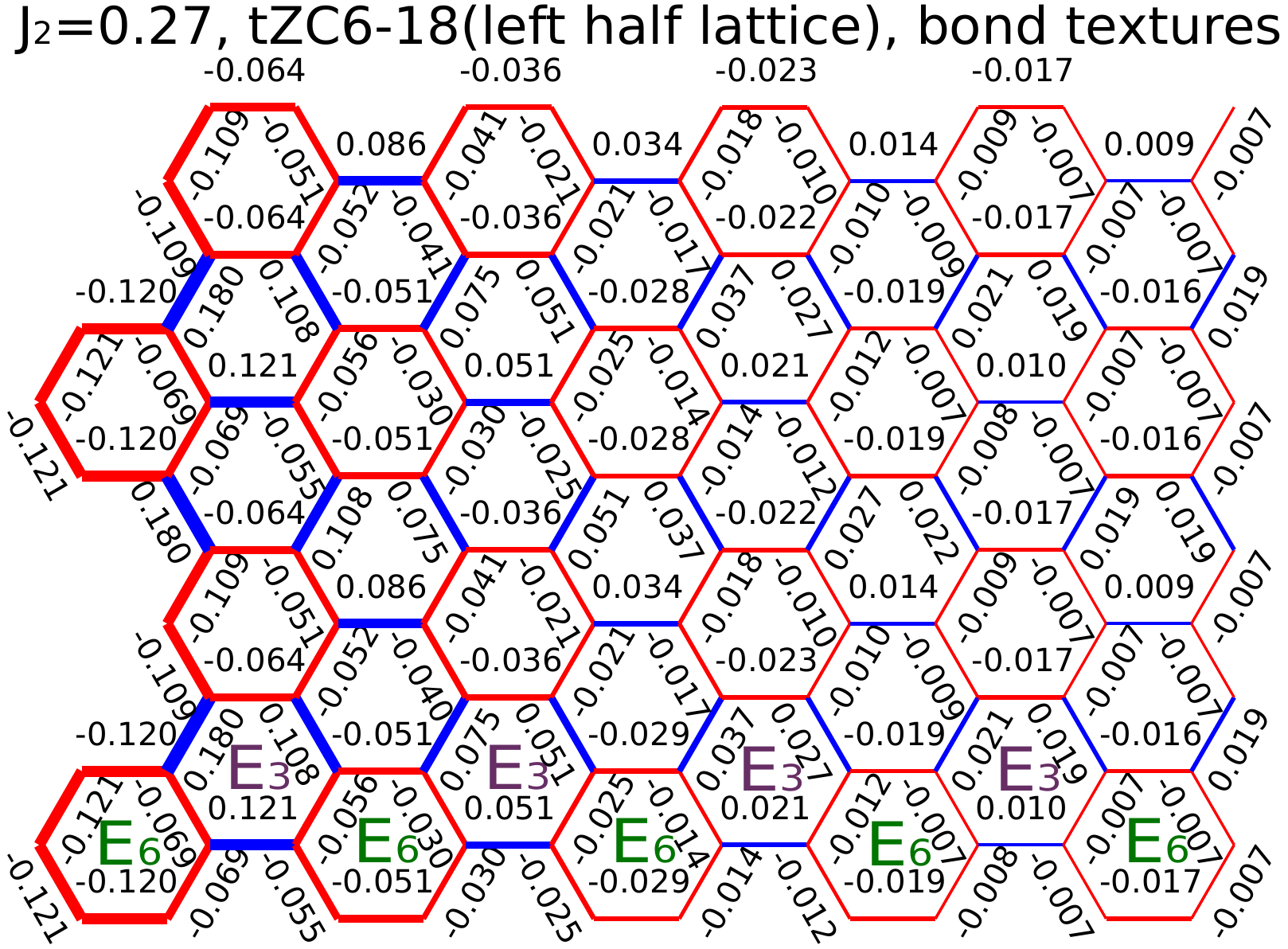}
\caption{The PVB bond texture $B_{i,j}$ for $J_2=0.27$ on a tZC6-18 lattice (left half of the lattice is shown). The summation of the bond textures on a hexagon with $6$ negative bond textures (red bonds) is denoted as $E_6$, while that on a hexagon with $3$ red bonds is $E_3$.  We define the PVB order parameter as the energy difference between two neighboring such hexagons, $P\equiv|E_6-E_3|$.} \label{tXC_PVB}
\end{figure}

On the tZC cylinder, the trimmed edges can select one of the three degenerate PVB states on the ZC cylinder.  Figure~\ref{tXC_PVB} shows the bond texture for $J_2=0.27$ on a tZC6-18 lattice.  The red hexagons with negative textures at the boundaries strongly pin the PVB state and induce the local PVB order. In the PVB state, the ``resonating'' hexagons have six negative bond textures $B_{i,j}$, while the other hexagons have three negative bonds. We define the summations of the bond textures on these two kinds of hexagons as $E_6$ and $E_3$, respectively (see Fig.~\ref{tXC_PVB}). Therefore, we can define the local PVB order parameter as the difference between two adjacent $E_6$ and $E_3$, i.e. $P\equiv|E_6-E_3|$. To measure the decay of the PVB order, we study the order parameter $P(d)$ along a row in the system (like the row with $E_6$ and $E_3$ symbols in Fig.~\ref{tXC_PVB}), where $d$ is the distance of the hexagons from boundary. We estimate $\xi_{P}$ by measuring the decay of $P(d)$ along $x$ direction from edge to bulk.

\begin{figure}[H]
\includegraphics[width = 1.0\linewidth,clip]{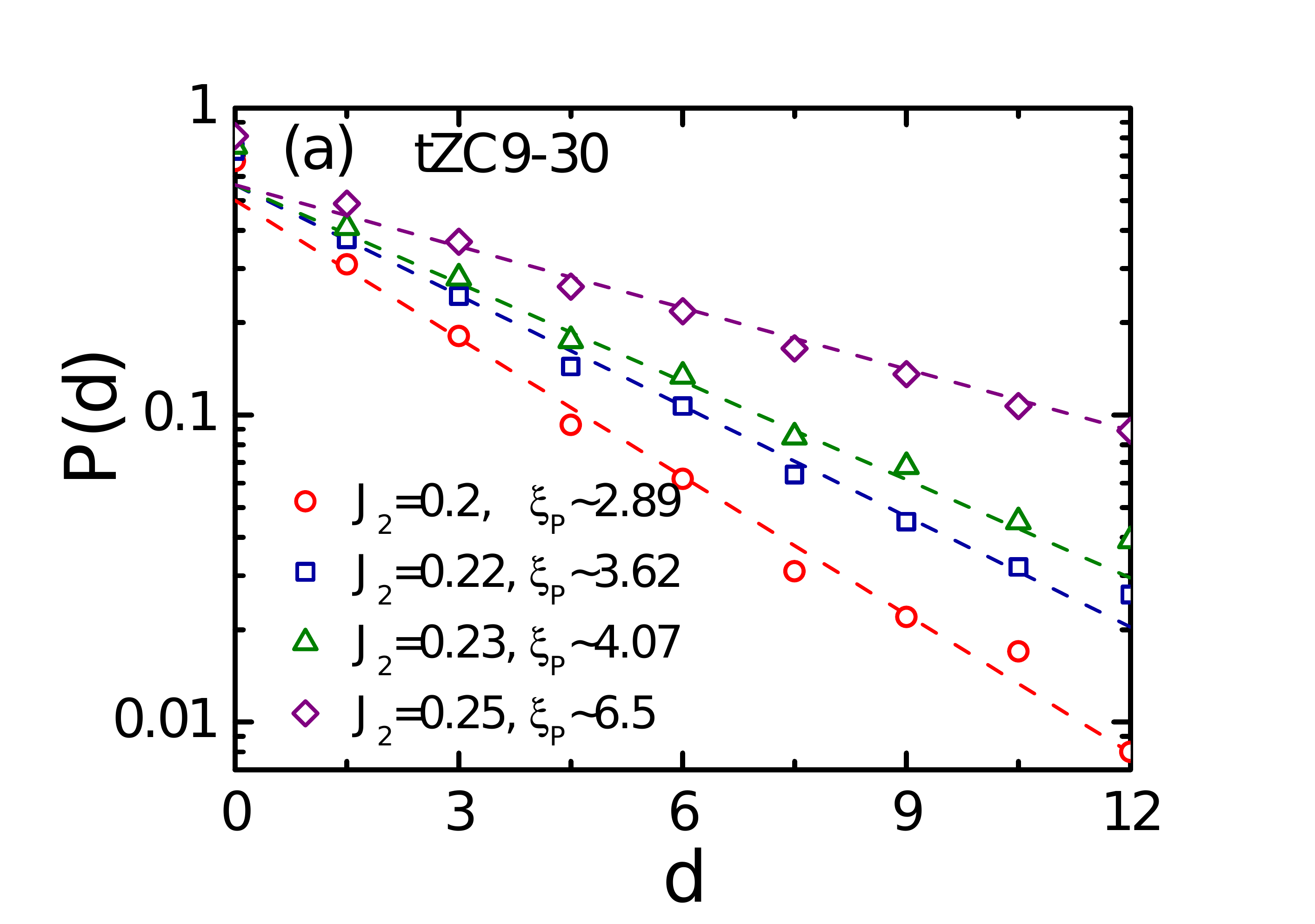}
\includegraphics[width = 1.0\linewidth,clip]{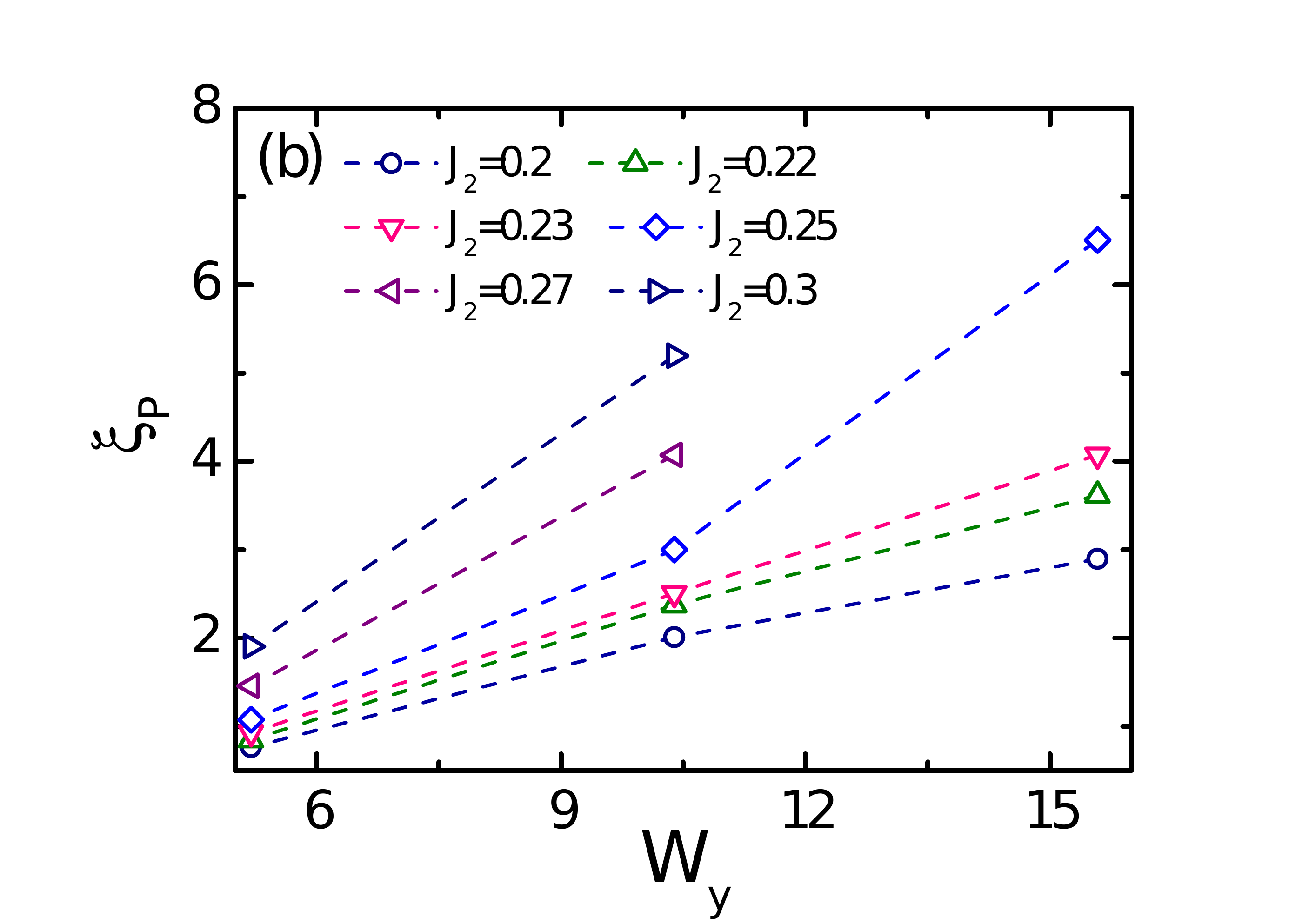}
\includegraphics[width = 1.0\linewidth,clip]{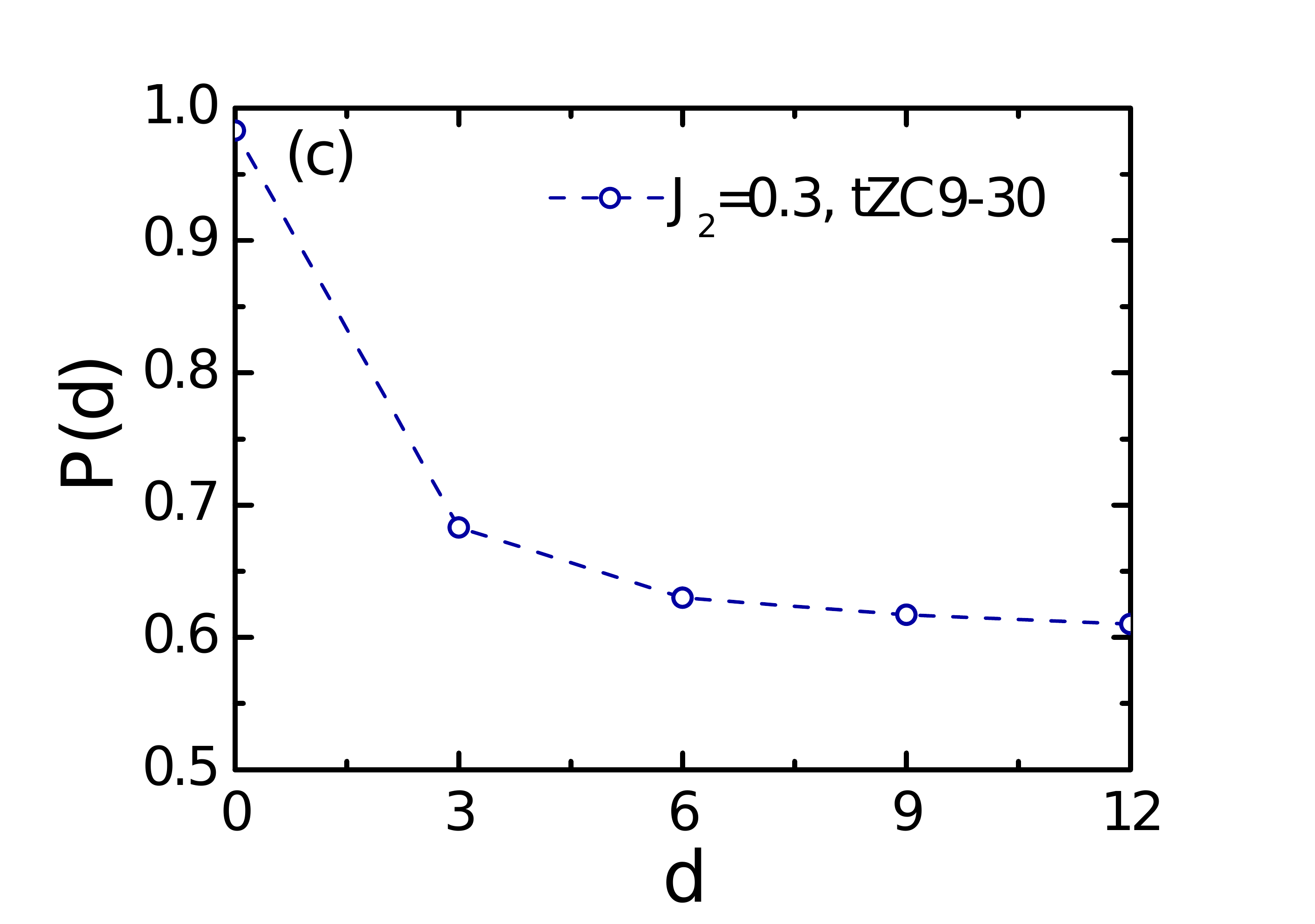}
\caption{(a) Log-linear plot of $P(d)$ on the tZC9-30 lattice for various $J_2$ couplings. (b) Circumference dependence of decay length $\xi_{P}$ on the tZC3-12, tZC6-18, and tZC9-30 cylinders for various $J_2$ couplings. For $J_2 = 0.27$ and $0.3$ on tZC9-30 cylinder, the systems have long-range PVB order and thus $\xi_{P}$ is divergent. (c) Real-space decay of $P(d)$ for $J_2=0.3$ on the tZC9-30 cylinder showing bulk PVB order.} \label{tXC_decay}
\end{figure}

The log-linear plot of $P(d)$ on the tZC9-30 cylinder for various $J_2$ is shown in Fig.~\ref{tXC_decay}(a). $P(d)$ decays exponentially and we can estimate $\xi_P$. In Fig.~\ref{tXC_decay}(b), we present the circumference dependence of $\xi_P$ on the tZC3-12, tZC6-18, and tZC9-30 cylinders for various $J_2$. For $J_2<0.25$, $\xi_P$ grows slower than the linear behavior with increasing width, while for $J_2\geq 0.25$, $\xi_P$ increases strongly. For $J_2=0.27$ and $0.3$, we find the long-range PVB order emerging on the tZC9-30 cylinder; Fig.~\ref{tXC_decay}(c) illustrates the non-zero PVB order in the bulk for $J_2=0.3$. The system appears to have the PVB order for $J_2 \gtrsim 0.25$ on the tZC cylinder, which is consistent with our observations on the AC cylinder.

By comparing the PVB decay length $\xi_P$ on the AC and tZC cylinders, we notice that the PVB order on the tZC cylinder grows faster than that on the AC cylinder on our studied finite sizes. The AC cylinder accommodates both the PVB and competing SVB orders, which might suppress the PVB order. On the other hand, the tZC cylinder frustrates the SVB order and at the same time provides strong seed for the PVB order at the edges, and this might enhance the PVB order throughout. Thus, the PVB order might also be overestimated on the tZC cylinder around $J_2=0.25$.

\subsection{PVB order on ZC cylinder}

\begin{figure}[tbp]
\includegraphics[width = 1.0\linewidth,clip]{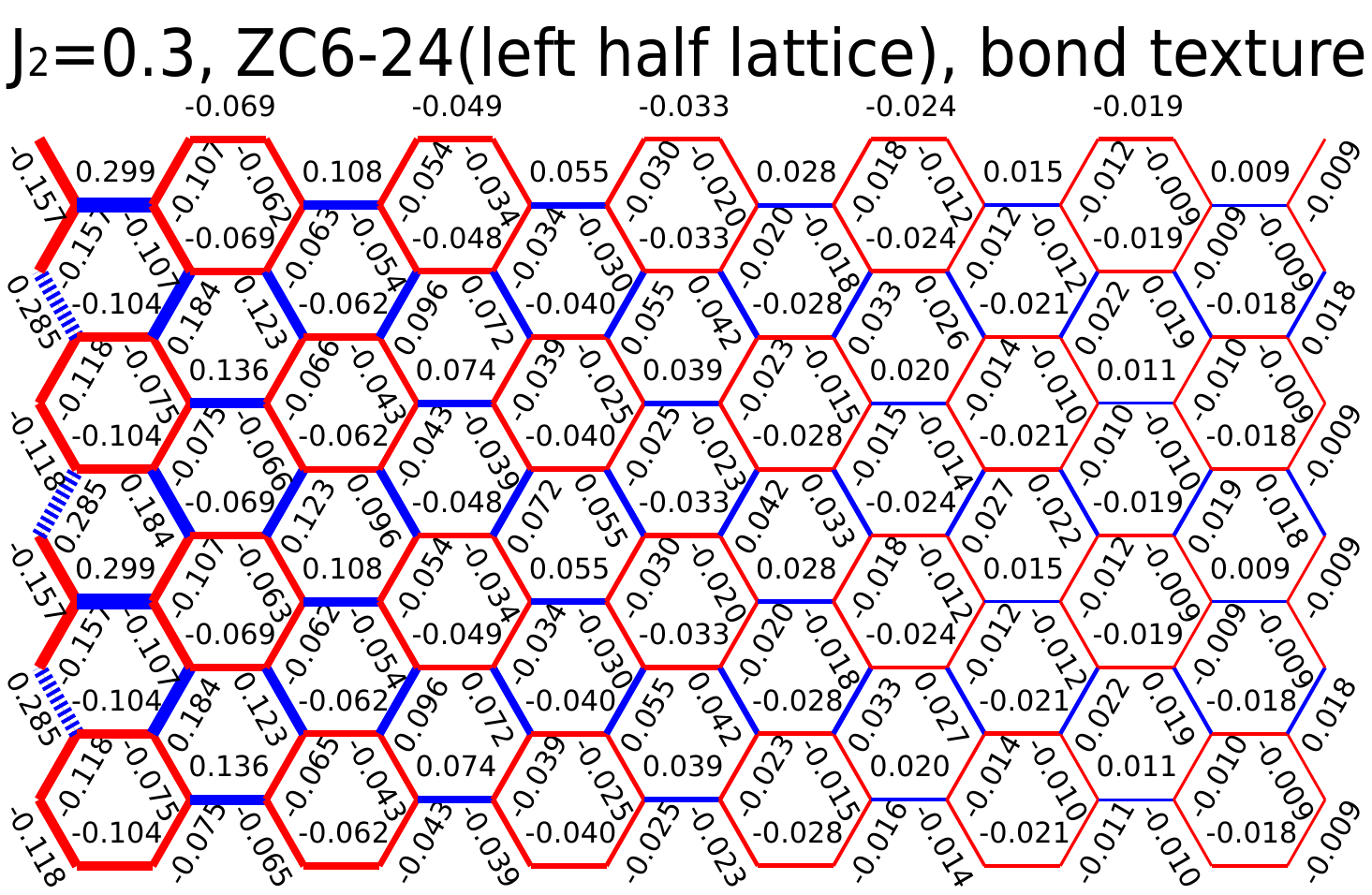}
\caption{The PVB bond texture $B_{i,j}$ for $J_2=0.3$ on the ZC6-24 cylinder with pinning appropriate for the PVB order. We show only the left half part of the lattice. The blue dashed lines indicate the bonds with pinning coupling $J_{\rm pin}=0.5$.}
\label{XC_texture}
\end{figure}

\begin{figure}[tbp]
\includegraphics[width = 1.0\linewidth,clip]{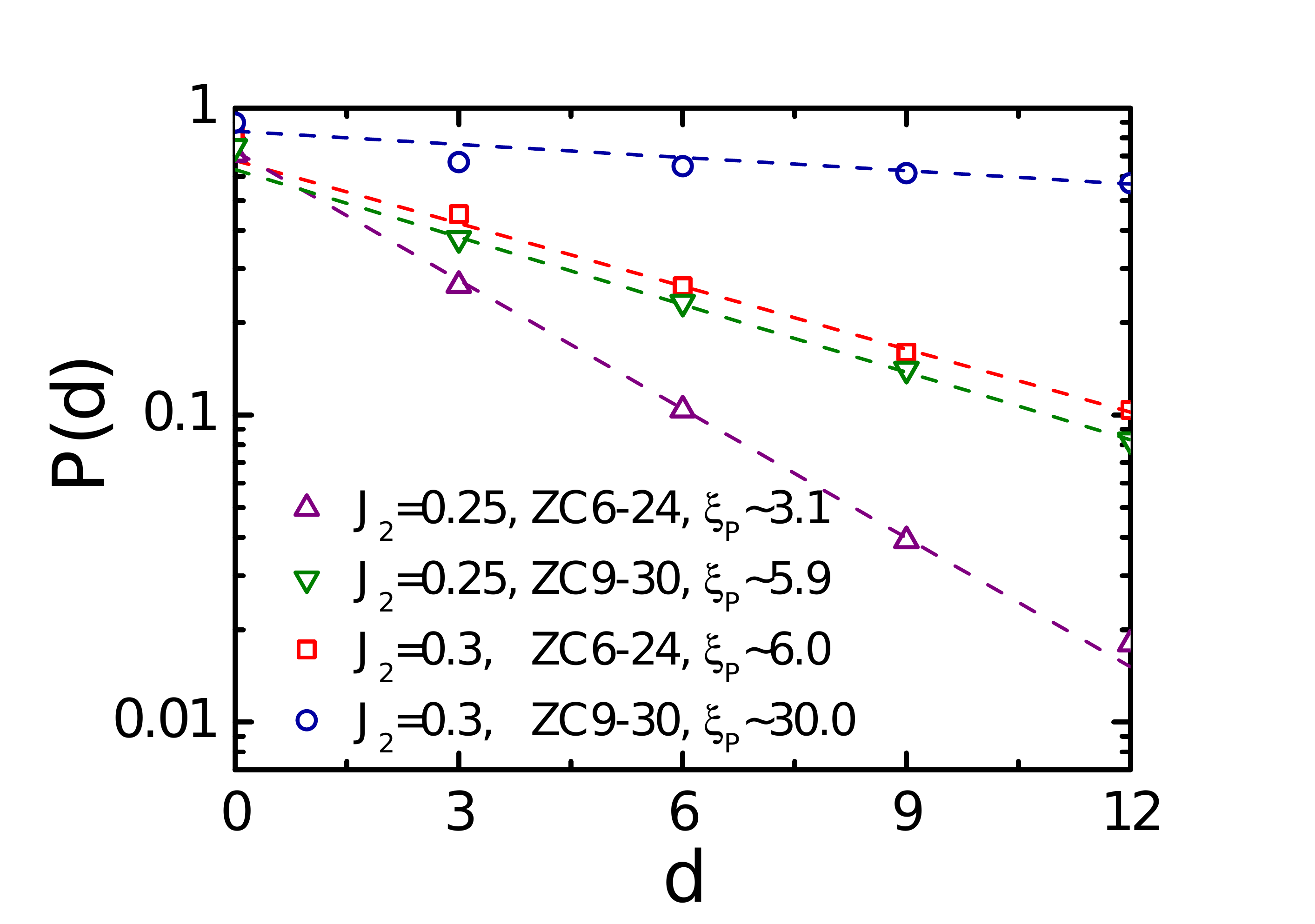}
\caption{Log-linear plot of PVB order on the ZC cylinder with PVB-pinning $J_{\rm pin}=0.5$ as in shown Fig.~\ref{XC_texture} for $J_2=0.25$ and $0.3$.}
\label{XC_decay}
\end{figure}

Finally, we summarize our data on cylinder with zigzag edges. To lift the degeneracy of the PVB state on the ZC cylinder, we can modify the bonds near the open boundaries to pin unique local PVB order. A simple way is to reduce the NN coupling to $J_{\rm pin}<J_1$ for selected bonds--namely, each in every three bonds--along the zigzag boundaries at both left and right ends of ZC cylinder.  Figure~\ref{XC_texture} shows the PVB bond texture for $J_2=0.3$ on the ZC6-24 cylinder where we reduced the NN coupling of the dashed blue bonds to $J_{\rm pin}=0.5$, which induces the local PVB order and selects unique PVB pattern in the system.

The PVB order parameter on ZC cylinder can be defined as that on tZC cylinder. In Fig.~\ref{XC_decay}, we present the log-linear plot of the PVB order on the ZC6-24 and ZC9-30 cylinders with $J_{\rm pin}=0.5$ for $J_2=0.25$ and $0.3$.  We study different ZC$m$-$n$ cylinders with ratio $n/m$ between $3$ and $4$ (the decay length is almost the same for fixed $m$).  For $J_2=0.25$, we find the decay lengths on the ZC6 ($\xi_P \simeq 3.1$) and ZC9 ($\xi_P \simeq 5.9$) cylinders consistent with those on the tZC cylinders in Fig.~\ref{tXC_decay}(b).  For $J_2=0.3$, $\xi_{P}\simeq 6.0$ for the ZC6-24 cylinder, close to $5.2$ on the tZC6 cylinder. On the ZC9-30 cylinder, we find that the obtained state is sensitive to the number of optimal states and sweep steps. By keeping about $16000$ states, we obtain a uniform state in the bulk of the system with the decay length a bit smaller than that of ZC6-24, but after keeping more than $20000$ states and increasing the number of sweeps, the ground-state energy is reduced and a strong PVB pattern emerges.

From the measurements of the width dependence of the PVB decay length $\xi_{P}$ on different cylinders with circumference as large as $W_y=15$, we find that the PVB order vanishes in the region $0.22 < J_2 \leq 0.25$, but grows strongly for $J_2 > 0.25$. Our observations of the PVB order are close to the DMRG results in Ref.~\onlinecite{PRL_110_127205}.

\section{Spin gap}

\begin{figure}[tbp]
\includegraphics[width = 1.0\linewidth,clip]{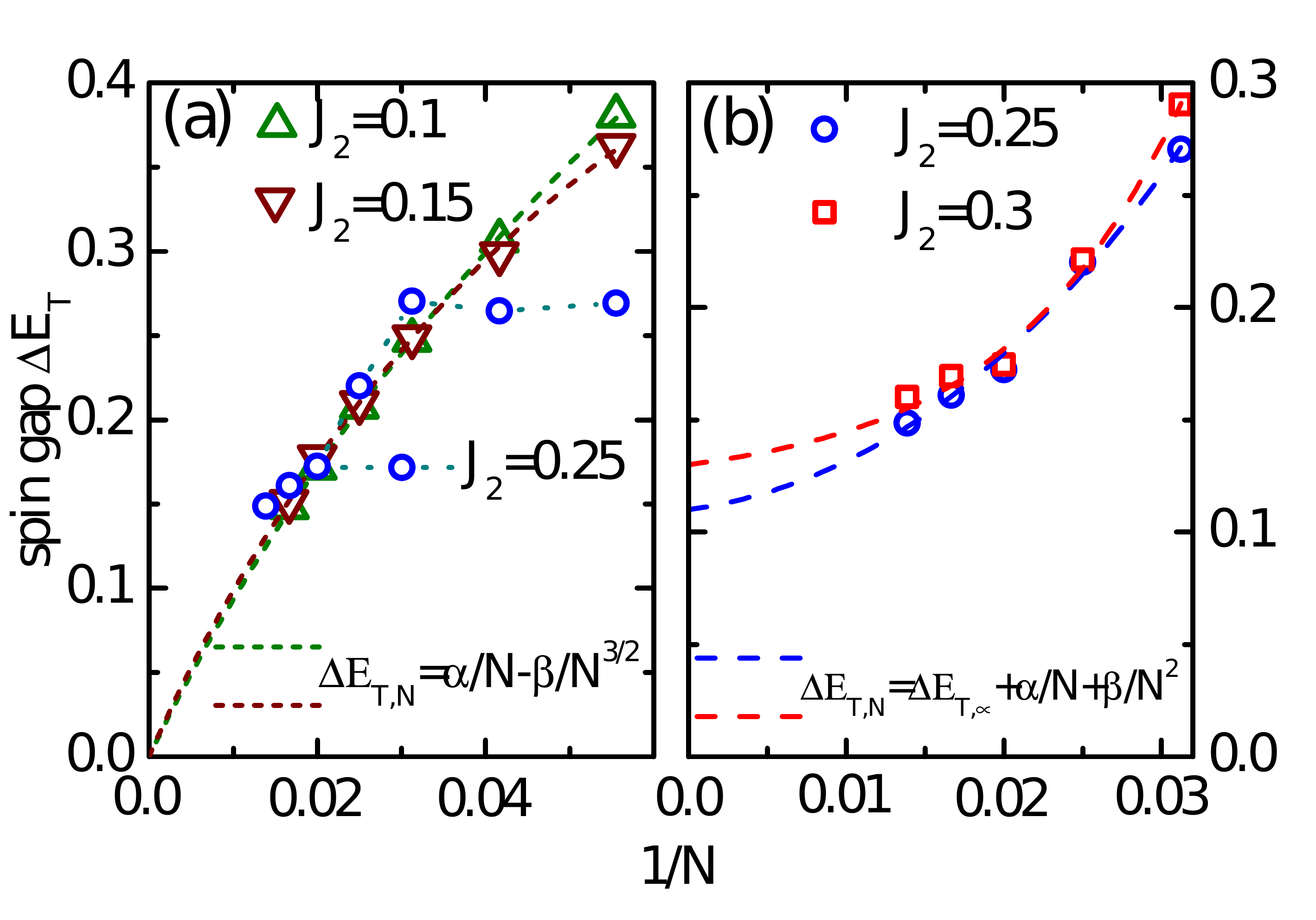}
\caption{Spin gap obtained from torus. (a) At $J_2=0.1$ and $0.15$, the spin gaps are extrapolated to zero as $\Delta E_{T,N}=\alpha/N-\beta/N^{3/2}$ from the samples $N=2\times 3\times 3$, $2\times 4\times 3$, $2\times 4\times 4$, $2\times 5\times 4$, $2\times 5\times 5$, and $2\times 6\times 5$. (b) At $J_2=0.25$ and $0.3$, the spin gaps are extrapolated to finite values as $\Delta E_{T,N}=\Delta E_{T,\infty}+\alpha/N+\beta/N^{2}$ from the larger samples $N=2\times 4\times 4$, $2\times 5\times 4$, $2\times 5\times 5$, $2\times 6\times 5$, and $2\times 6\times 6$.}\label{gap}
\end{figure}

In the N\'{e}el phase with broken SU(2) symmetry, we have gapless Goldstone modes, and consequently the spin gap should vanish, while in the PVB phase the spin gap appears due to the broken translational symmetry. Spin gap has been studied by U(1) DMRG in fully open system\cite{PRL_110_127203} and cylinder system,\cite{PRL_110_127205} both of which find the non-zero spin gap in the intermediate coupling regime. Here we study the spin gap on torus system, which is free from the edge excitations in the open boundary.

Figure~\ref{gap}(a) shows the finite-size spin gaps for $J_2=0.1$ and $0.15$ for different torus sizes from $N=2\times 3\times 3$ to $2\times 6\times 5$ with 2D-like clusters. The finite size scaling shows that these data can be extrapolated to zero quite well using the first two terms in the $1/\sqrt{N}$ expansion $\Delta E_{T,N}=\alpha/N-\beta/N^{3/2}+\mathcal{O}(1/N^{2})$,\cite{EPJB_20_241,PRB_39_2608} which is expected for the N\'{e}el phase.

For $J_2=0.25$, the size dependence of the gap changes substantially. The gaps at smaller $N$ are near constant which could be consistent with a lattice-symmetry-broken state; however, at larger sizes ($N=32$ to $N=72$), they drop with $N$ but have a trend of saturating toward a finite value. In Fig.~\ref{gap}(b), we fit the finite-size gaps from larger system sizes by the formula $\Delta E_{T,N}=\Delta E_{T,\infty}+\alpha/N+\beta/N^2$, and find non-zero spin gap in the thermodynamic limit for $J_2=0.25$ and $0.3$. The finite spin gap at $J_2=0.3$ is consistent with the PVB order. For $0.22<J_2<0.25$, it is hard to identify the size of the spin gap from the extrapolations of our finite-size data, which suggests either a small or vanishing gap.

The above DMRG results show that the magnetic and PVB orders are vanished for $0.22<J_2\leq 0.25$, which could be consistent with the observation of a spin liquid. For a gapped SL in this region, our DMRG measurements would suggest a continuous transition from the N\'{e}el to gapped SL.\cite{PRL_110_127203} Although there are some new theories suggesting such a transition,\cite{PRB_82_024419,PRB_86_214414} the conventional viewpoint is that the collinear N\'{e}el order is not connected to SL through continuous transition in 2D system.\cite{IJMP_5_219} On the other hand, a very recent Quantum Monte Carlo study\cite{arxiv_1302_1408} of a honeycomb J-Q model found a continuous transition from the N\'{e}el to the PVB phase and proposed a ``deconfined quantum criticality'' scenario (although in general such a transition could also be discontinuous).  Therefore, our proposal of spin liquid in this region can be challenged by other possibilities like the deconfined quantum criticality with long correlation length.

\section{Entanglement entropy}

\begin{figure}[tbp]
\includegraphics[width = 1.0\linewidth,clip]{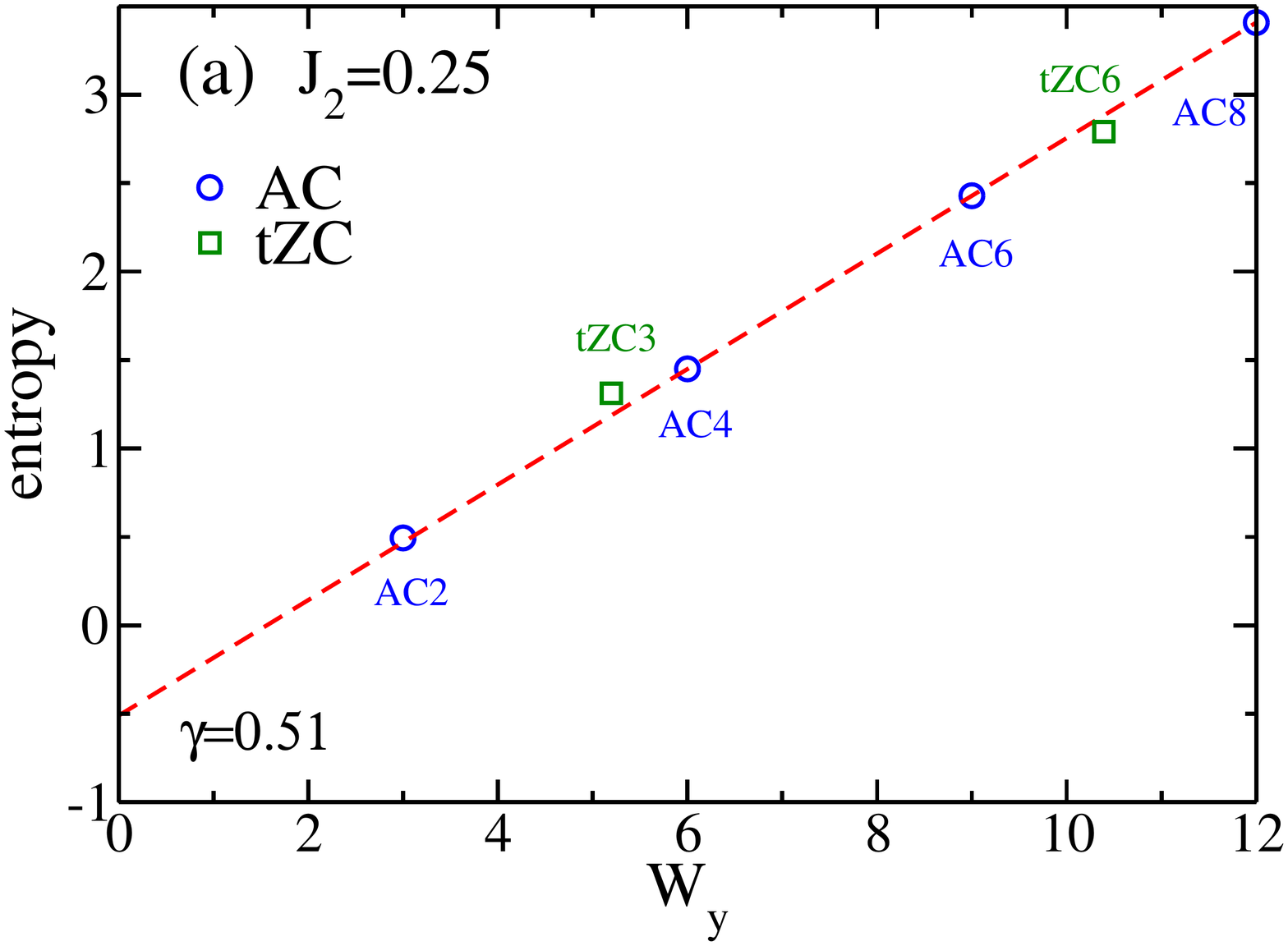}
\includegraphics[width = 1.0\linewidth,clip]{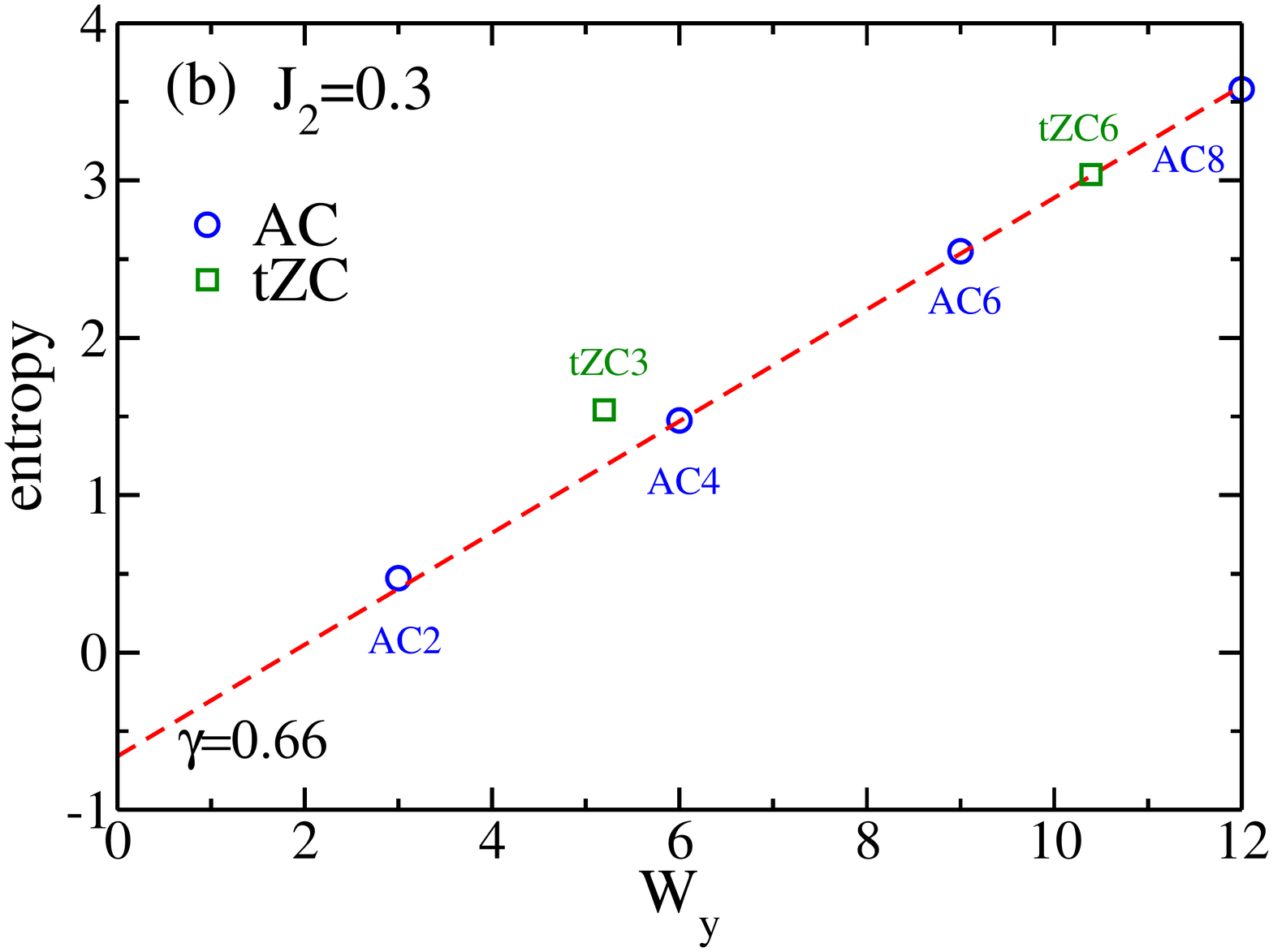}
\caption{Circumference dependence of the entanglement entropy in large $L_1$ limit measured on both AC and tZC cylinders for (a) $J_2 = 0.25$, and (b) $J_2 = 0.3$. The linear extrapolations of the entanglement entropy using data on both cylinders lead to the topological entanglement entropy $\gamma=0.51$ and $0.66$, respectively.}\label{ee}
\end{figure}

For a gapped quantum state with topological order, the topological entanglement entropy $\gamma$ is proposed to characterize the non-local feature of entanglement.\cite{PRL_96_110404,PRL_96_110405}  The Renyi entropies of a subsystem $A$ with density matrix $\rho_A$ are defined as $S_n = (1-n)^{-1}\ln({\rm Tr}\rho_A^{n})$, and the Von Neuman entropy is defined as $n\rightarrow 1$ limit of the Renyi entropy.  For such a state with topological order, the Renyi entropies have the form $S_n = \alpha L - \gamma$, where $L$ is the boundary of the subsystem and all other terms vanish in large $L$ limit. Here $\alpha$ is a non-universal constant, while a positive $\gamma$ term is a correction to the area law of entanglement and reaches a universal value determined by the total quantum dimension $D$ of the quasiparticle excitations of the state.\cite{PRL_96_110404,PRL_96_110405}

To establish the nature of the ground state as a possible topologically nontrivial SL state, positive evidences are highly desired, particularly for such a non-Braivais lattice system where a trivial insulator may exist without  breaking any symmetry.\cite{EPL_93_37007,Bravis1} Recently, a number of topologically ordered states have been identified from the TEE by extrapolating the EE of the minimum entropy state (MES) on long cylinders through DMRG calculations.\cite{NP_8_902} It is suggested that this method should be efficient when all the correlation lengths are short compared with cylinder width, and the DMRG would favor the MES on long cylinders in this situation.\cite{NP_8_902}

We obtain the EE on cylinders by making a cut for subsystems in the middle of lattice along the vertical direction. We scale the EE to large $L_1$ limit for each circumference on both AC and tZC cylinders to obtain the EE of the possible MES, and plot the circumference dependence of the resulting Von Neumann entropy to extrapolate the TEE.

For $0.22<J_2\lesssim 0.25$, the system appears to have no dimer order, and the spin correlation lengths are short on the studied sizes. By extrapolating the EE, we obtain $\gamma=0.51$ in this region. As presented in Fig.~\ref{ee}(a) for $J_2=0.25$, the best linear fit of the EE using data on both the AC and tZC cylinders gives $\gamma=0.51$. If the 2D system is magnetically disordered in this region, the non-zero $\gamma$ could indicate nontrivial topological feature. Somewhat surprisingly, for larger $J_2$ in the PVB phase, we obtain $\gamma$ close to $\ln 2$, which is the TEE value of $Z_2$ SL. As shown in Fig.~\ref{ee}(b) for $J_2=0.3$, the best linear fit of the EE using data on both the AC and tZC cylinders gives $\gamma=0.66$. A possible explanation could be that on the system sizes in Fig.~\ref{ee}(b), the long-range PVB order does not emerge. Thus, the wavefunction in the bulk of the lattice might appear like a gapped SL, which could lead to a TEE close to $Z_2$ SL. However, as the long-range PVB order sets in for larger sizes, the scaling of the TEE may graduate change, which cannot be directly checked due to our simulation limit:  In DMRG calculations, we need to keep more states to converge the EE compared to ground-state energy,\cite{PRB_79_205112} and for $W_y>12$ our calculations of the EE are likely not fully converged.  While our entropy data is well converged for $W_y \leq 12$, TEE is still geometry dependent as we can see from comparing the AC and tZC cylinders (e.g., if we used only the data on the tZC cylinders, the linear extrapolations in Fig.~\ref{ee} would give $\gamma$ values close to zero).

\section{Comparisons with variational Monte Carlo}
\label{sec:VMC}

\begin{figure}[t]
\includegraphics[width = 1.0\linewidth,clip]{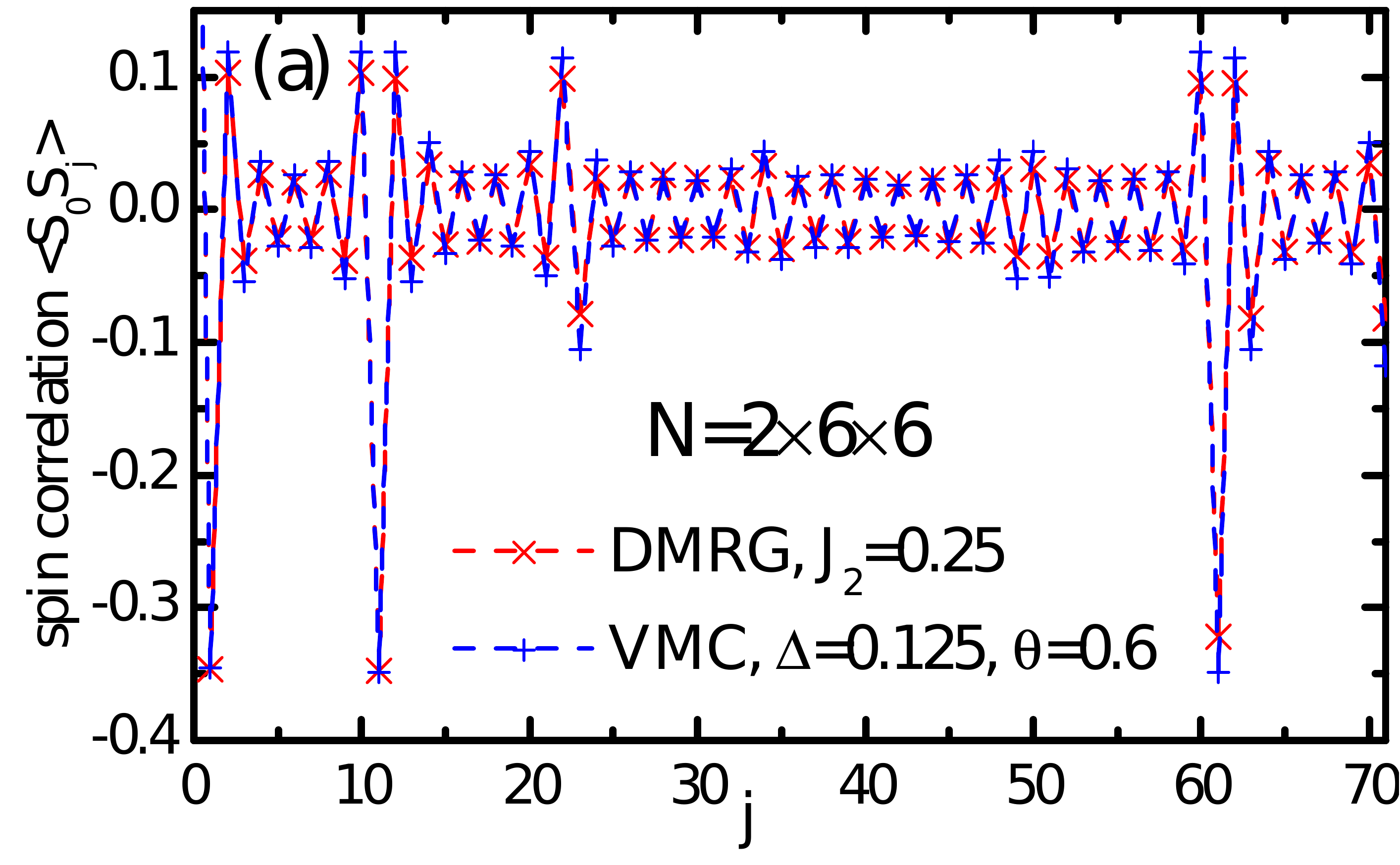}
\includegraphics[width = 1.0\linewidth,clip]{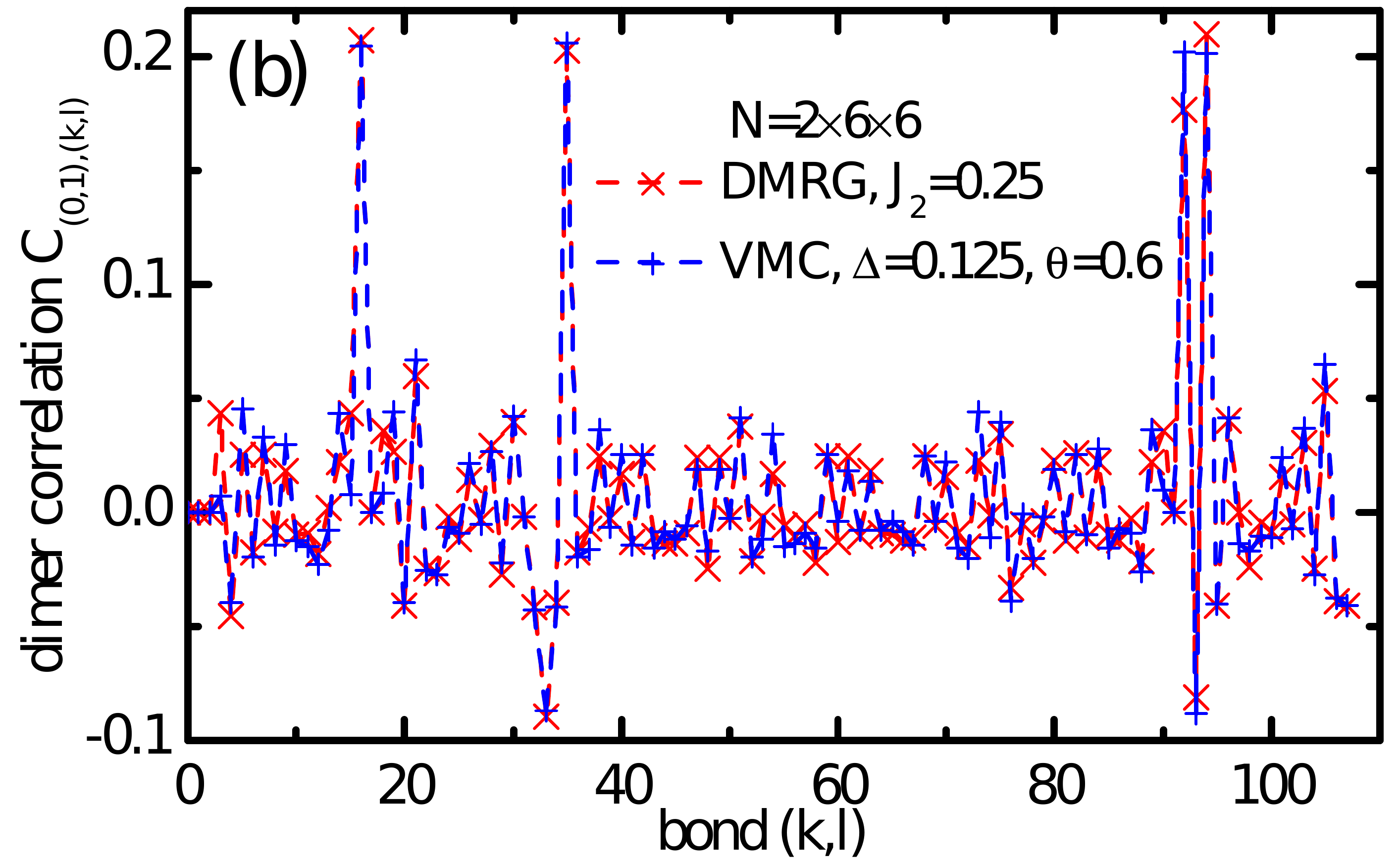}
\caption{Comparisons of DMRG and VMC results on torus for (a) spin and (b) dimer correlations at $J_2 = 0.25$ and $N = 2\times 6\times 6$. The VMC wave function is the SPS state of Ref.~\onlinecite{PRB_84_024420} with $\Delta = 0.125$ and $\theta = 0.6$ and represents a $Z_2$ spin liquid state (VMC results look similar for a range of $\theta$ including $\theta=0$).
Sites $j$ and bonds $(k,l)$ are ordered in a typewriter fashion going first in the $\vec{a}_2$ direction in Fig.~\ref{phase}.
}\label{VMC_DMRG}
\end{figure}

As a test of possible SL for $0.22 < J_2 \leq 0.25$, we compare the DMRG results against VMC calculations using so-called Sublattice Pairing State (SPS)\cite{PRB_84_024420,PRL_107_087204} constructed as follows.
We use slave fermion representation of spins, ${\bf S}_i = \frac{1}{2} f_{i\alpha}^\dagger {\bm \sigma}_{\alpha\beta} f_{i\beta}$, with the constraint of precisely one spinon per site.  We consider spinon mean field with hopping and pairing\cite{PRB_65_165113}
\begin{eqnarray}
H_{\rm mf} &=& -\sum_{ij, \alpha} t_{ij} f_{i\alpha}^\dagger f_{j\alpha} + \sum_{ij} \left(\Delta_{ij} f_{i\up}^\dagger f_{j\dn}^\dagger + \Hc \right) \notag \\
&-&\sum_{i, \alpha} \mu_i f_{i\alpha}^\dagger f_{i\alpha},
\end{eqnarray}
with $t_{ji} = t_{ij}^*$ and $\Delta_{ji} = \Delta_{ij}$.  The SPS state has real-valued nearest-neighbor hopping $t$ and complex-valued second-neighbor pairing with a specific pattern of phases: $\Delta_{ij} = |\Delta| e^{i\theta}$ for $i,j$ from one sublattice of the honeycomb lattice and $\Delta_{ij} = |\Delta| e^{-i\theta}$ for $i,j$ from the other sublattice.  There are two variational parameters, $|\Delta|/t$ and $\theta$.  We set $\mu_i = 0$, which in the SPS state automatically gives one spinon per site on average.  We find the mean field ground state and then perform Gutzwiller projection into the physical spin Hilbert space and use this as our trial wave function.  We calculate the energies and various correlation functions using standard VMC techniques.\cite{Ceperley1977, Gros89}

Our VMC energetics study finds that for $J_2 \leq 0.2$ the optimal state is essentially a ``Dirac spin liquid'' with $|\Delta| \approx 0$.  For $J_2 \geq 0.2$ the optimal $|\Delta|$ rises continuously, and our onset of non-zero $\Delta$ is different from the result in Ref.~\onlinecite{PRL_107_087204}.  Our optimal $\theta$ tends to remain near zero, although we find that the energetics is not very sensitive to $\theta$ in a range of values.  For example, for $J_2 = 0.25$ we find that the energy is minimized at $|\Delta| = 0.125, \theta \approx 0$, but with a nearly flat dependence on $\theta \in (0, 0.7)$.

In Figs.~\ref{VMC_DMRG}(a) and \ref{VMC_DMRG}(b), we compare the spin and dimer correlations in the DMRG ground state at $J_2 = 0.25$ and in the VMC state with $|\Delta| = 0.125$ and $\theta = 0.6$ on the $N = 2\times 6 \times 6$ torus sample.  The spinon mean field has antiperiodic boundary conditions in both directions, which gives the lowest trial energy in this sample.  Strictly speaking, this trial state breaks lattice rotation symmetry because of the boundary conditions; however, we found that the anisotropy in bond energies is only few percent and essentially does not affect the comparisons with the DMRG.  The agreement between the DMRG and VMC is striking.  Even though the wave function represents a gapped $Z_2$ spin liquid with no magnetic or dimer order on long distances, the spin correlations in the VMC are a bit stronger than in the DMRG, and the same is true about the dimer correlations.

The above results suggest that despite fairly strong such correlations in our DMRG measurements, they are reasonable for a gapped $Z_2$ spin liquid on such finite samples.  We have chosen to present $\theta = 0.6$ to emphasize this point, but results for a range of $\theta$ including $\theta = 0$ look very similar on this size (we do not see significant difference between correlation functions for $\theta = 0$ and $\theta=0.6$ even up to size $2\times 15 \times 15$ that we studied in VMC).  However, we expect a qualitative difference between $\theta \neq 0$ and $\theta = 0$ SPS states on long distances.\cite{PRB_84_024420}  Specifically, even though the spinon dispersion has a gap for all $\theta$, the gauge structure is $Z_2$ only when $\theta \neq 0$, while the gauge structure is U(1) when $\theta = 0$ (i.e., this case is equivalent to a pure hopping ansatz).  As we further discuss in Appendix~\ref{app:SBVMC}, we expect the U(1) ansatz to be unstable beyond mean field and can view its appearance as suggesting proximitity to a Valence Bond Solid.  In Appendix~\ref{app:SBVMC}, we also present VMC energetics using Schwinger Boson (SB) wave functions; while the SB study is limited to only small sizes, we find general agreement with the slave fermion VMC and similar hints of proximity to a U(1) regime and VBS order.  It would be interesting to extend the present VMC work to include true VBS order directly in the wave functions and to try to match with the DMRG results on open cylinders.

\section{Summary and discussion}

\begin{table*}
\begin{tabular}{|c|c|c|}
\hline
$J_2$ coupling & Earlier results & Our results \\
\hline
$0 \leq J_2 < 0.22$ & N\'{e}el phase established by ED\cite{EPJB_20_241,JPCM_23_226006,PRB_84_024406}, DMRG\cite{PRL_110_127203,PRL_110_127205}, \textit{et.al.} & N\'{e}el phase \\
\hline
$0.22 \leq J_2 \leq 0.26$ & Controversy among N\'{e}el (DMRG\cite{PRL_110_127205}), PVB (DMRG\cite{PRL_110_127203}) and SL (VMC\cite{PRL_107_087204}) & possible SL \\
\hline
$0.26 < J_2 \leq 0.35$ & Controversy between PVB (DMRG\cite{PRL_110_127203,PRL_110_127205}) and SL (Mean-field\cite{PRB_82_024419,PRB_84_024420,PRB_87_024415}, VMC\cite{PRL_107_087204}) & PVB phase \\
\hline
\end{tabular}
\caption{Earlier results in each interval of $J_2$ coupling of $J_1$-$J_2$ honeycomb model, as well as the results established by our works. We have found a N\'{e}el phase for $J_2 < 0.22$, a PVB phase for $0.26 < J_2 \leq 0.35$. For $0.22 \leq J_2 \leq 0.26$, we find a possible SL in the system.} \label{backgroundI}
\end{table*}

\begin{table*}
\begin{tabular}{|c|c|c|c|}
\hline
$J_2$ coupling & Ref. \onlinecite{PRL_110_127203} & Ref. \onlinecite{PRL_110_127205} & Our results \\
\hline
$0 \leq J_2 < 0.22$ & N\'{e}el phase & N\'{e}el phase & N\'{e}el phase \\
\hline
$0.22 \leq J_2 \leq 0.26$ & \textit{PVB phase} & \textit{N\'{e}el phase} & \textit{possible SL} \\
\hline
$0.26 < J_2 \leq 0.35$ & PVB phase & PVB phase & PVB phase\\
\hline
\end{tabular}
\caption{The ground states of $J_1$-$J_2$ honeycomb model from recent DMRG works. All these works have found a N\'{e}el phase for $0 \leq J_2 < 0.22$ and a PVB phase for $0.26 < J_2 \leq 0.35$. For $0.22 \leq J_2 \leq 0.26$, a PVB\cite{PRL_110_127203} and a N\'{e}el\cite{PRL_110_127205} phase have been proposed. In our work, we find that the system is a SL in this region, or has a deconfined quantum critical point from N\'{e}el to PVB phase at $J_2 \simeq 0.26$.} \label{backgroundII}
\end{table*}

In summary, we have studied the phase diagram of the spin-$1/2$ $J_1$-$J_2$ Heisenberg model on honeycomb lattice by means of DMRG with SU(2) symmetry and VMC. By implementing SU(2) symmetry in DMRG, we can study cylinder geometry with circumference slightly over $W_y=15$ and torus with size up to $2\times 6\times 6$. We compute the square of the staggered magnetic moment, $m_{s}^{2}$, on both torus and ZC cylinder (from ZC4-8 to ZC9-18). By extrapolating the finite-size $m_{s}^{2}$ to the thermodynamic limit, we estimate that the N\'{e}el order vanishes at $J_2\simeq 0.22$.

In order to investigate the PVB order in the intermediate region, we first study the dimer-dimer correlation functions and the dimer structure factor on torus up to the size $2\times 6\times 6$. We observe two weak peaks of the dimer structure factor at $\textbf{q}=(2\pi/3,4\pi/3)$ and $(4\pi/3,2\pi/3)$, indicating the PVB pattern of the dimer correlations. The absence of peak at $\textbf{q}=0$ indicates the vanishing SVB order. We study cylinders to determine the PVB order on larger sizes. For a system with even weak dimer order in the 2D limit, the PVB order decay length from open edge to bulk in cylinder geometry is found to grow faster than linear with increasing width, and will diverge on large size.\cite{PRB_85_134407} Therefore, we study the width dependence of the PVB order decay length on the AC, tZC, and ZC cylinders for various $J_2$ couplings. We estimate the decay length of the PVB order $\xi_P$ by fitting the exponential decay of PVB order parameter from boundary to bulk. We find that for $J_2\lesssim 0.25$, $\xi_P$ grows slowly and appears to saturate in the 2D limit. For $J_2>0.25$, $\xi_P$ grows strongly with increasing width, implying a possible PVB state in the 2D limit.

We also study the spin gap on torus in both the N\'{e}el and the intermediate regions. For $J_2=0.1$ and $0.15$ in the N\'{e}el phase, the finite-size spin gaps are extrapolated to zero as $\Delta E_{T,N}=\alpha/N-\beta/N^{3/2}$, which is the expected behavior for the N\'{e}el state. For $J_2\gtrsim 0.25$, the spin gaps extrapolate to finite values, which are consistent with the observed PVB order.

We expect that a gapped $Z_2$ SL has a non-zero TEE. We study the EE on both AC and tZC cylinders and extrapolate the EE in the large $L_1$ limit to obtain the TEE of the possible MES. We find the TEE value of $\gamma=0.51$ for $0.22\lesssim J_2\lesssim 0.25$. For $J_2=0.3$, the TEE extrapolation gives $\gamma=0.66$, which is close to the TEE value of $Z_2$ SL, $\ln 2$. However, since for this $J_2$ we observe the PVB order on larger sizes, the obtained $\ln 2$ value from our range of system sizes may not represent a signature of topological order in the thermodynamic limit; instead, it could still be due to the finite-size effect.

As a test of a possible SL for $0.22 < J_2 \leq 0.25$, we also study this region by VMC simulations and directly compare the spin and dimer correlation functions from DMRG and VMC results on torus. For $J_2=0.25$ on the $2\times 6\times 6$ torus, we find the striking match of the DMRG results with the VMC wave function of a $Z_2$ SL. The match of correlation functions further indicates that the ground states on such finite-size sample for $0.22 < J_2 \leq 0.25$ are consistent with a $Z_2$ SL. However, the optimal VMC states are close to a gapped U(1) SL point of the SPS ansatz, which may render it unstable towards a Valence Bond Solid, and the VMC is not conclusive about the ultimate state on long distances.
From our DMRG data, the possibility of spin liquid also competes with an alternative scenario of a quantum critical point between the N\'{e}el and PVB phases, which we are unable to exclude with our finite-size studies.

In Tables \ref{backgroundI} and \ref{backgroundII}, we show earlier and recent DMRG results of $J_1$-$J_2$ honeycomb model, which have controversies for $0.22 \leq J_2 \leq 0.35$. In our DMRG calculations, we have found N\'{e}el order for $J_2 < 0.22$ and solid evidences of a weak PVB order for $0.26 < J_2 \leq 0.35$. In the interesting region $0.22 \leq J_2 \leq 0.26$, we exclude the PVB order clearly by large-scale results, which indicates a SL, or a N\'{e}el phase.\cite{PRL_110_127205} 

In our search for robust spin liquid regimes, we have also performed studies of the honeycomb $J_1$-$J_2$-$J_3$ model with ferromagnetic $J_3$, complementary to the work in Ref.~\onlinecite{PRB_84_024406} which studied antiferromagnetic $J_3$.  However, we find that the Staggered Valence Bond solid becomes very prominent already for small ferromagnetic $J_3$, leaving only a very small possible SL regime. We quickly find a direct N\'{e}el to SVB transition, which moves to smaller $J_2$ values upon increasing $|J_3|$. It would be interesting to look for other modifications of the model that could provide robust spin liquid on the honeycomb lattice.

\acknowledgments
We would like to thank Z.-Y.~Zhu, S.~White, and D.~Huse for extensive discussions of their related work.  We also acknowledge stimulating discussions with  L.~Balents, B.~Clark, H.-C.~Jiang, A.~Vishwanath, C.-K.~Xu, and Z.-C.~Gu. S.S.G thanks J.-Z.~Zhao for help in developing DMRG code. This research is supported by the National Science Foundation through grants DMR-0906816 (S.S.G. and D.N.S.), DMR-1206096 (O.I.M.), DMR-1101912 (M.P.A.F.), and by the Caltech Institute of Quantum Information and Matter, an NSF Physics Frontiers Center with support of the Gordon and Betty Moore Foundation (O.I.M. and M.P.A.F.).

\appendix

\section{Summary of VMC energetics with Schwinger boson wave functions}
\label{app:SBVMC}
We have also considered projected Schwinger boson (SB) wave functions using so-called Zero Flux (ZF) state of Ref.~\onlinecite{PRB_82_024419}.  Here we use slave boson representation of spins,
${\bf S}_i = \frac{1}{2} b_{i\alpha}^\dagger {\bm \sigma}_{\alpha\beta} b_{i\beta}$,
with the constraint of one slave boson per site.  The Schwinger boson mean field Hamiltonian is
\begin{equation}
H_{\rm SB, mf} = \sum_{ij} \left( A_{ij} b_{i\up}^\dagger b_{j\dn}^\dagger + \Hc \right) - \mu \sum_{i, \alpha} b_{i\alpha}^\dagger b_{i\alpha} ~.
\end{equation}
For simplicity, we only include SB ``pairing'' terms $A_{ij}$, which are expected to be appropriate for the antiferromagnetic spin interactions.\cite{PRB_45_12377, PRB_74_174423, PRB_82_024419}  We require $A_{ji} = -A_{ij}$ to satisfy SU(2) spin invariance.  The ZF ansatz has nearest-neighbor $A_{< ij >} = A_1$ for orientations $i \to j$ from one sublattice of the honeycomb lattice to the other.  It also has second-neighbor $A_{<< ij >>} = A_2$ for orientations going clockwise (counter-clockwise) around up (down) triangles formed by the second-neighbor bonds inside each hexagon, see Fig.~3 in Ref.~\onlinecite{PRB_82_024419}.  We find the mean field ground state and then perform Gutzwiller projection into the physical Hilbert space as described in Ref.~\onlinecite{PRB_84_020404}.  The result is a Resonating Valence Bond (RVB) wave function with specific singlet amplitudes determined from the SB ansatz.  Using direct permanent calculations in the $S^z$ basis,\cite{PRB_84_020404} we can perform measurements for such wave functions on systems with up to $N = 50$ sites.  

The Zero Flux state has two variational parameters, $A_2$ and $\mu$ (setting $A_1 = 1$).  When $\mu$ is very close to the bottom of the band---i.e., the Schwinger bosons are very close to condensation---the RVB singlet amplitudes are power-law long-ranged and the wave function is a good approximation to the N\'{e}el state.\cite{Beach, PRB_84_020404} On the other hand, when $\mu$ is away from the bottom of the band, the wave function represents a short-range RVB state.

\begin{figure}[t]
\centerline{\includegraphics[width=\columnwidth]{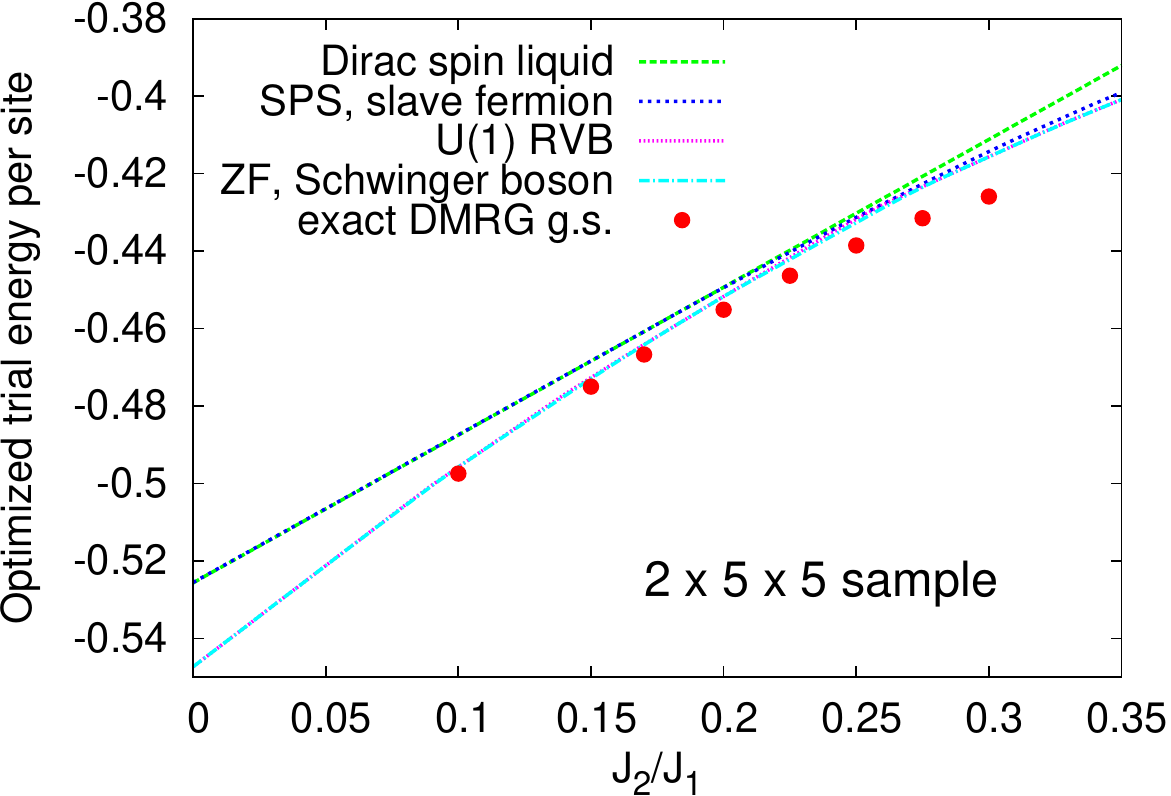}}
\caption{
Variational energies on the $2 \times 5 \times 5$ torus comparing ZF Schwinger boson and SPS slave fermion ansatze, together with the exact DMRG ground state energy.  The Dirac spin liquid is obtained by setting $\Delta = 0$ in the SPS state.  The U(1) RVB state is obtained by setting $A_2 = 0$ in the ZF state. The ZF state can also realize a long-ranged RVB state and can thus provide a good approximation to the N\'{e}el state for $J_2 < 0.2$. In the putative spin liquid region for larger $J_2$, the optimal ZF state is a short-ranged RVB state whose energetics is very similar to the optimal SPS case. Note that the optimal parameters in both the ZF and SPS cases are close to U(1) regime in the respective ansatze as explained in the text. The DMRG on larger clusters is indispensible in determining the ultimate nature of the ground state.
}
\label{fig:en5x5x2}
\end{figure}

Figure~\ref{fig:en5x5x2} shows optimized trial energies for the ZF Schwinger boson wave function and the SPS slave fermion wave function, on a $2 \times 5 \times 5$ system, together with the exact DMRG results.  The optimal parameters in the SPS are similar to the ones discussed in Sec.~\ref{sec:VMC}.  Here we focus on the ZF SB case.  For small $J_2 < 0.2$, the optimized chemical potential is close to the bottom of the spinon band, and the SB wave function provides an accurate approximation to the N\'{e}el ordered state.  For larger $J_2$, the chemical potential moves far below the bottom of the band, and the wave function represents a short-range RVB liquid.  We find that the optimal $A_2$ is small in this regime, $A_2/A_1 \lesssim 0.1$.  We illustrate this in Fig.~\ref{fig:en5x5x2} by plotting also the trial energy with fixed $A_2 = 0$ and varying only the chemical potential, which gives essentially the optimal ZF SB energy.

In the absence of $A_2$, the resulting RVB state has only singlets connecting the different sublattices.  This is usually viewed as a U(1) spin liquid, hence the label ``U(1) RVB'' in the figure.  The common belief is that the U(1) spin liquid with gapped spinons is unstable beyond mean field in (2+1)d once gauge fluctuations are included and that the ultimate state is Valence Bond Solid.\cite{PRL_66_1773, PRB_42_4568, PRB_65_165113} Since the constructed formal wave function does not include the gauge fluctuations and is not the full theory, it need not represent a qualitatively accurate physical ground state and need not have such a VBS order. Numerical studies of U(1) RVB wave functions on the square lattice found exponenentially decaying spin correlations but power-law decaying dimer correlations.\cite{PRB_82_180408, PRB_84_174427} Therefore, while they are not accurate representations of the VBS phase, we can still view the U(1) RVB wave functions as suggesting incipient VBS order.  Because of this, it would be interesting to determine long-distance properties of the U(1) RVB wave functions also on the honeycomb lattice (this was not possible with our method using permanents but should be feasible with valence bond Monte Carlo as in Refs.~\onlinecite{PRB_82_180408, PRB_84_174427}).

On the other hand, if we had a substantial non-zero $A_2$, the resulting state would be a stable $Z_2$ spin liquid.  As the variational results stand, they are not conclusive about the robustness of the spin liquid state and can also be interpreted as suggesting proximity to a VBS order.  It is ultimately for unbiased numerical studies like the DMRG to determine the true nature of the ground state.

We conclude by noting that Ref.~\onlinecite{PRB_84_024420} conjectured that the ZF Schwinger boson wave function and the SPS slave fermion wave function represent the same $Z_2$ spin liquid.  On a crude level, Fig.~\ref{fig:en5x5x2} shows that the optimized energetics is very similar in the two states.  We have also compared the spin and dimer structure factors in the optimized Schwinger boson and slave fermion states and found that they are quantitatively close.  This supports the conjecture in Ref.~\onlinecite{PRB_84_024420}, but we caution that both wave functions are close to the U(1) regime.  We have also compared such properties of the ZF and SPS wave functions deep in the presumed $Z_2$ regime and found them to be similar.  Note, however, that we have compared only correlations of local observables and only on relatively small $N \leq 50$ clusters, while it is important to compare topological properties\cite{Grover2013} to ascertain that the two states are in the same phase; we leave this as an interesting open problem.


\begin{thebibliography}{99}
\bibitem{nonBravais} Interestingly, for the non-Bravais honeycomb lattice, it is also possible to have a trivial insulator that does not break any symmetry but has no topological order.\cite{EPL_93_37007,Bravis1}
\end{thebibliography}
\end{document}